\documentclass[10pt,journal,compsoc]{IEEEtran} 

\IEEEoverridecommandlockouts

\usepackage{bbding}

\usepackage[most]{tcolorbox}
\usepackage{xcolor}     
\usepackage{enumitem}

\usepackage{adjustbox}
\usepackage{pifont} 
\usepackage{colortbl} 
\usepackage{booktabs} 
\usepackage{array} 

\usepackage{amsmath,amsfonts}
\usepackage{algorithmic}
\usepackage{algorithm}
\usepackage{array}
\usepackage[caption=false,font=normalsize,labelfont=sf,textfont=sf]{subfig}
\usepackage{textcomp}
\usepackage{stfloats}
\usepackage{url}
\usepackage{verbatim}
\usepackage{graphicx}
\usepackage{cite}
\usepackage{adjustbox}

\usepackage{booktabs}
\usepackage{multirow} 
\usepackage{xcolor}
\usepackage{amsfonts,amssymb}
\usepackage{amsthm} 
\usepackage{wasysym}

\usepackage{multirow}

\usepackage[utf8]{inputenc}
\usepackage{amsmath,amssymb,amsfonts}

\usepackage{graphicx}
\usepackage[edges]{forest}
\usepackage{xcolor} 
\definecolor{customgreen}{HTML}{70AD47} 
\definecolor{customorange}{HTML}{E6990E} 

\definecolor{myblue}{RGB}{112, 156, 156}
\definecolor{lightblue}{RGB}{237, 250, 252}

\definecolor{myyellow}{HTML}{fffaea}
\definecolor{mycyan}{HTML}{edf8fc}
\definecolor{mygrey}{HTML}{f0f1f7}

\usepackage[colorlinks=true, linkcolor=blue, urlcolor=blue, citecolor=blue]{hyperref}
\usepackage{url}

\begin{document}



\title{A Survey of Foundation Model-Powered Recommender Systems:
From Feature-Based, Generative to Agentic Paradigms}

\author{Chengkai Huang, Hongtao Huang, Tong Yu, Kaige Xie, Junda Wu, Shuai Zhang, Julian Mcauley, Dietmar Jannach and Lina Yao
\IEEEcompsocitemizethanks{
\IEEEcompsocthanksitem C. Huang is with the School of Computer Science and Engineering, The University of New South Wales. E-mail: chengkai.huang1@unsw.edu.au.
\IEEEcompsocthanksitem H. Huang is with the School of Computer Science and Engineering, The University of New South Wales. E-mail: hongtao.huang@unsw.edu.au
\IEEEcompsocthanksitem T. Yu, is with Adobe Research. E-mail: tyu@adobe.com 
\IEEEcompsocthanksitem  K. Xie is with the School of Computer Science, Georgia Institute of Technology. E-mail: kaigexie@gatech.edu. 
\IEEEcompsocthanksitem J. Wu is with the University of California, San Diego. E-mail: juw069@ucsd.edu.
\IEEEcompsocthanksitem S. Zhang is with ETH Zurich. E-mail: cheungshuai@outlook.com. 
\IEEEcompsocthanksitem J. Mcauley is with the University of California, San Diego. E-mail: jmcauley@eng.ucsd.edu.
\IEEEcompsocthanksitem D. Jannach is with the University of Klagenfurt. E-mail: Dietmar.Jannach@aau.at.
\IEEEcompsocthanksitem L. Yao is with The University of New South Wales and CSIRO’s Data61. E-mail:  lina.yao@data61.csiro.au.

}
\thanks{(Corresponding authors: Chengkai Huang.)}
}


 

\markboth{IEEE TRANSACTIONS ON KNOWLEDGE AND DATA ENGINEERING, SUBMISSION 2025}%
{Shell \MakeLowercase{\textit{et al.}}: Bare Demo of IEEEtran.cls for Computer Society Journals}

\IEEEtitleabstractindextext{%
\begin{abstract} 


Recommender systems (RS) have become essential in filtering information and personalizing content for users. RS techniques have traditionally relied on modeling interactions between users and items as well as the features of content using models specific to each task. The emergence of foundation models (FMs), large scale models trained on vast amounts of data such as GPT, LLaMA and CLIP, is reshaping the recommendation paradigm.
This survey provides a comprehensive overview of the Foundation Models for Recommender Systems (FM4RecSys), covering their integration in three paradigms: 
(1) \emph{Feature-Based} augmentation of representations, 
(2) \emph{Generative recommendation} approaches, and 
(3) \emph{Agentic interactive systems}. 
We first review the data foundations of RS, from traditional explicit or implicit feedback to multimodal content sources. We then introduce FMs and their capabilities for representation learning, natural language understanding, and multi-modal reasoning in RS contexts. 
The core of the survey discusses how FMs enhance RS under the feature-based paradigm (improving feature representations), the generative paradigm (directly generating recommendations or related content), and the agentic paradigm (enabling autonomous recommendation agents and simulators). 
Afterward, we examine FM applications in various recommendation tasks: Top-N recommendation, sequential recommendation, zero/few-shot scenarios, conversational recommendation, and novel item/content generation. Through an analysis of recent research, we highlight key opportunities that have been realized (e.g., improved generalization, better explanations, reasoning ability) as well as challenges encountered (e.g., cross-domain generality, interpretability, fairness, and multimodal integration). Finally, we outline open research directions and technical challenges for next-generation FM4RecSys, such as multimodal recommender agents, retrieval-augmented frameworks, lifelong learning for long user sequences, efficiency and cost issues, etc. 
This survey not only reviews the state-of-the-art methods but also provides a critical analysis of the trade-offs among the feature-based, the generative, and the agentic paradigms, outlining key open issues and future research directions.

\end{abstract}

}

\maketitle
\IEEEpeerreviewmaketitle

\newcommand{\kai}[1]{{\color{blue} [kai: #1]}}

\section{Introduction}



\IEEEPARstart{R}ecommender Systems (RSs) have become critical in a wide range
of domains, from e-commerce and social media to healthcare and education
\cite{RicciRS15,ZhangYST19}. They aim to deliver personalized content by capturing
user preferences, item characteristics, and contextual signals. Over the past
decade, the field has witnessed remarkable progress, driven by advancements in
deep learning architectures and the increasing availability of large-scale user
behavior data. Despite these achievements, traditional RSs still face persistent challenges in capturing subtle user preferences, handling cold-start scenarios, and
providing transparent, context-rich explanations. 
These challenges limit the effectiveness of purely domain-specific or small-scale models in providing accurate and diverse recommendations.

In parallel, Foundation Models (FMs) have made significant strides in areas such as natural language processing, computer vision, and multi-modal tasks \cite{li2024multimodal}. Recently, FMs have been reshaping recommender system architectures---boosting performance, enabling novel modes of user interaction, and demonstrating strong potential in capturing complex user-item relationships while generalizing across a broader spectrum of recommendation tasks.
To be specific, Foundation Models for Recommender Systems (\textbf{FM4RecSys}) refer to leveraging the knowledge from pre-training and recommendation datasets to capture rich representations of user preferences, item features, and contextual variables for improving personalization and prediction accuracy in recommendation tasks. 
Meanwhile, Foundation Models (FMs)---large-scale, pre-trained models with strong task generalization capabilities that provide a unified and flexible modeling paradigm for various downstream recommendation tasks \cite{survey1}.
Unlike conventional methods that rely on meticulously engineered features or narrow architectures, FMs leverage broad pre-training over massive corpora, enabling stronger generalization and the ability to incorporate a wide variety of signals (text, images, audio, knowledge graphs, etc.). This flexibility can yield richer user/item representations and help overcome the data sparsity and cold-start issues that plague traditional collaborative filtering. 
Beyond boosting predictive accuracy, foundation models (FMs) unlock novel capabilities---including natural language explanations, interactive conversational interfaces, and even agentic decision-making. 
In particular, agentic frameworks leverage FMs to autonomously plan, reason, and adapt within dynamic environments by incorporating iterative user feedback and real-time contextual understanding.
Next, we dive into the motivations behind existing works that incorporate foundation models into recommender systems, aiming to deepen our understanding of how FMs are applied and what impact they have across different recommendation tasks.

\subsection{Motivation}

We enumerate the primary motivations driving the research in the evolving landscape of FM4RecSys, aiming to provide a comprehensive understanding of the factors that contribute to the development and adoption of FM-powered RSs.

\textbf{Enhanced Generalization Capabilities.} Foundation Models are designed to learn from large-scale data, enabling them to understand complex patterns. FMs can generalize better to new, unseen data \cite{foudnation_survey8}. In the context of RSs, this means that FMs can more accurately predict user preferences and behaviors, especially in scenarios with sparse data or novel items (defined as zero-shot/few-shot recommendations in some papers  \cite{gao2023chat,zeroshotRec,hou2024zero}).
Through zero-shot/few-shot inference of user preferences and item attributes, FMs are able to deliver effective recommendations, even in the absence of extensive interaction history.


\textbf{Elevated Recommendation Experience.}  Foundation Models foster a transformative interface paradigm for recommendation systems, significantly altering the user interaction experience. For instance, conversational RS is a classic use case, the previous Conversational  (CRSs) \cite{GaoLHRC21,Lei0MWHKC20} predominantly rely on pre-established dialogue templates, a dependency that often constrains the breadth and adaptability of user engagements. In contrast, FMs introduce a paradigm shift towards more dynamic and unstructured conversational interactions, offering enhanced interactivity and flexibility. The interactive design allows for more engaging and natural user interactions with the system. Users can conversationally communicate their preferences, ask questions, and receive customized recommendations.

\textbf{Improved Explanation and Reasoning Capabilities.} Foundation Models
augment the explanation and reasoning capabilities of RS. Whereas traditional recommender
systems predominantly derive explanations from rudimentary sources such as user reviews
or elementary user behaviors, including co-purchased items or peer purchases, these
explanations are often bereft of in-depth logic and context \cite{LiZC20}. In contrast, Foundation Models possess the ability to formulate explanations that are
enriched with a comprehensive grasp of commonsense and user-specific context \cite{reasoning_survey}.
These models leverage an array of data, encompassing user preferences,
historical interactions, and distinctive item characteristics, to generate
explanations that are both more coherent and logically sound. Utilizing Foundation
Models to deeply interpret user behavior sequences and interests can significantly enhance the effectiveness of future recommender systems in complex scenarios \cite{Yanreasoning}, promising to advance informed and responsible decision-making
processes in areas like medicine and healthcare, e.g., treatment and diagnosis recommendations.


Given these benefits, a wave of research has begun exploring FM4RecSys. 
As traditional RSs struggle with issues such as data sparsity and rigid feature extraction, the emergence of FMs promises broader generalization capabilities. 
However, realizing this potential in real-world applications brings new challenges, such as real-time adaptation, computational efficiency, and interoperability that remain underexplored. To better understand both the opportunities and limitations, we provide a comprehensive and critical assessment of FM4RecSys, organized around three core paradigms and a range of recommendation tasks.


\subsection{Paradigms of FM-Powered Recommender Systems}\label{sub:paradigms}

\begin{figure}
    \centering
    \includegraphics[width=1\linewidth]{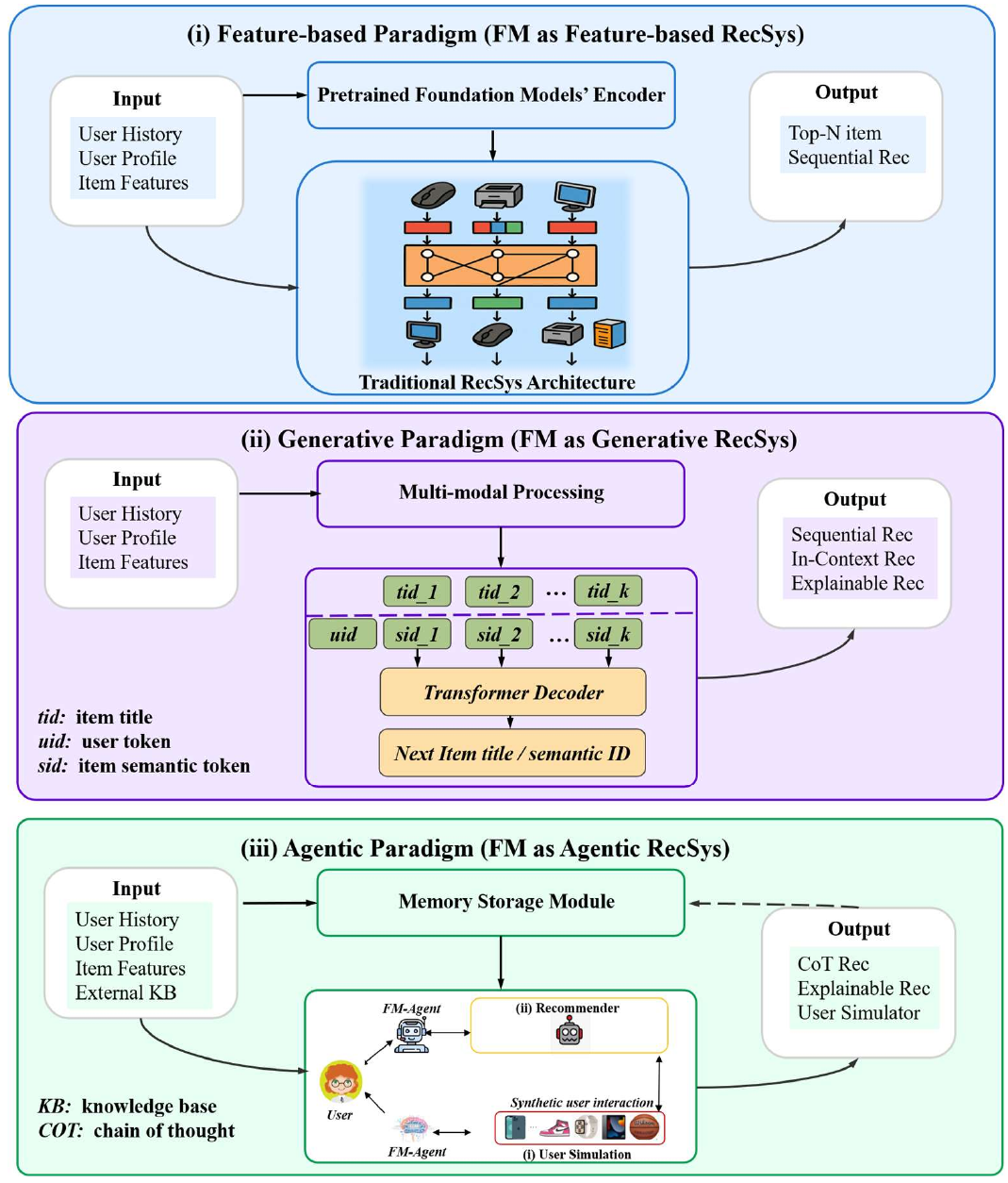}
    \caption{Three Paradigms of FM-Powered Recommender Systems}
    \label{fig:framework}
\end{figure}

How can FMs be integrated into recommender systems? We identify three paradigms of integration in current research: Feature-Based, Generative, and Agentic. These paradigms differ in the role the foundation model plays in the RS pipeline (from a passive feature provider to an active decision-maker). Figure \ref{fig:framework} provides a high-level comparison of these paradigms with examples.

\textbf{Feature-Based Paradigm:}
This approach treats foundation models as feature extractors to generate high-quality
embeddings for users, items, or interactions. For example, text-based FMs (e.g., BERT) \cite{DevlinCLT19} encode item descriptions or user reviews into semantic vectors, while vision-language models (e.g., CLIP) \cite{RadfordKHRGASAM21} align multimodal features (text, images) for
cross-domain recommendations. While effective, this paradigm often limits FMs to
auxiliary roles, decoupled from the core recommendation logic.

\textbf{Generative Paradigm:} This approach leverages the generative capabilities of FMs (e.g., GPT), this paradigm directly synthesizes recommendations as generated outputs \cite{rajput2023recommender}.
Examples include generating personalized explanations \cite{prompting_survey}, creating virtual items (e.g., advertising slogans, product designs), or predicting user preferences through autoregressive token prediction. However, such systems face challenges in controllability and alignment with user intent, as generation may prioritize fluency over relevance.

\textbf{Agentic Paradigm:} The emerging agentic paradigm reimagines RSs as autonomous agents powered by FMs \cite{huang2025towards}. These agents dynamically interact with users
(e.g., via natural language), reason about long-term preferences, and even take actions
(e.g., probing questions, multi-step planning) to refine recommendations. Unlike static models, agentic systems exhibit goal-driven behavior, leveraging tools (e.g., search engines, databases) and feedback loops to adapt to evolving contexts.

While feature-based and generative paradigms have advanced recommendation accuracy and diversity, the agentic paradigm represents a transformative shift toward proactive, explainable, and human-aligned systems. By integrating reasoning, tool use, and multi-turn interaction, agentic FMs address critical limitations of traditional RS: 
(i) Dynamic Adaptation: Agents continuously update user profiles based on real-time feedback, mitigating the cold-start and data sparsity issues. 
(ii) Multimodal Contextualization: They unify text, voice, and visual inputs to
capture nuanced preferences (e.g., interpreting a user’s screenshot of a product).
(iii) Ethical Alignment: To balance personalization with fairness and transparency, recent studies have explored constitutional AI techniques~\cite{bai2022constitutional}, which guide model behaviors by incorporating predefined ethical principles or human-aligned rules into the generation process. The rapid progress of LLM-based agents (e.g., AutoGPT\footnote{\url{https://github.com/microsoft/autogen}}, Meta’s CICERO \cite{meta2022human}) and retrieval-augmented generation (RAG) frameworks further demonstrates the feasibility of this paradigm.


\subsection{Distinguishing Features from Recent LLM-based RS Surveys}\label{sec_difference}

Research on FM4RecSys is accelerating, and several surveys have recently
reviewed parts of this emerging intersection.
Liu \textit{et al.} \cite{survey2} delved into the training strategies and
learning objectives of language modeling paradigm adaptations for recommenders, while Wu \textit{et al.} \cite{survey3} provided insights from both discriminative and generative viewpoints on Language Model-based Recommender Systems (LLM4Rec). Lin \textit{et al.} \cite{survey1} introduces two orthogonal perspectives: where and how to adapt LLMs in recommender systems. Fan \textit{et al.} \cite{survey4} offered an overview of LLMs for recommender systems, concentrating on paradigms such as pre-training, fine-tuning, and prompting. Lin \textit{et al.} \cite{survey5} summarized the current progress in generative recommendations, organizing them across various recommendation tasks.

\begin{figure*}[ht]
    \centering
    \includegraphics[width=0.95\textwidth]{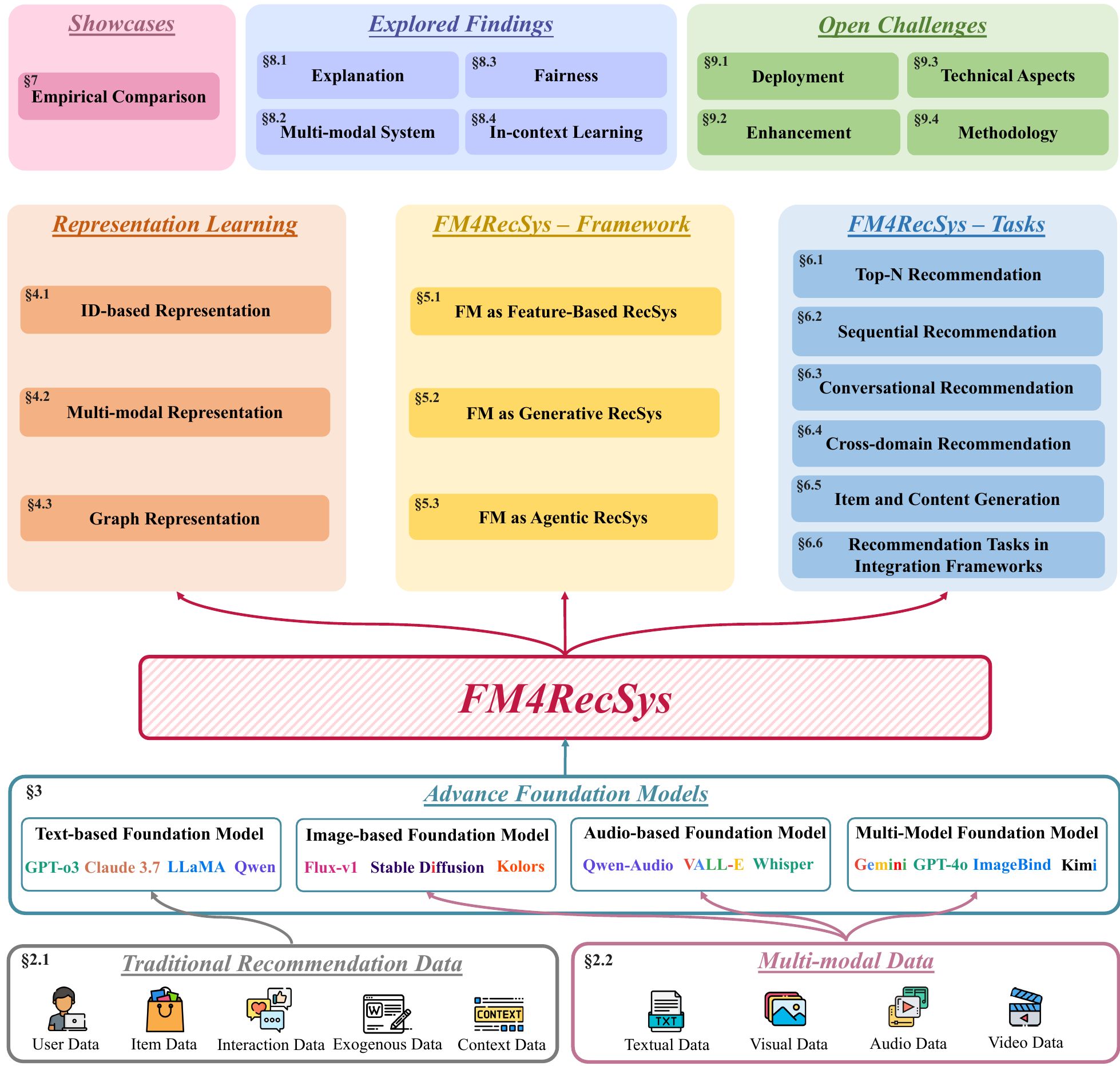}
    \caption{The taxonomy of FM4RecSys from data characteristics to open problems and opportunities. In contrast to previous surveys, our methodology introduces a unique viewpoint for examining the intersection of FM4RecSys from data characteristics to open problems and opportunities, which are detailed in Section \ref{sec_difference}.}
    \label{fig:overview}
\end{figure*}

{\textbf{Differences and Key Contributions:}} In contrast to previous surveys,
our survey takes a broader view: we cover foundation models beyond just LLMs,
including vision and multimodal models, and structure the discussion along a new
taxonomy spanning data, integration paradigms, tasks, and open challenges. In particular,
we emphasize a three-paradigm framework --- feature-based, generative, and agentic
--- for understanding how FMs can be leveraged in recommender systems, alongside a
range of downstream recommendation tasks that have been tackled with FMs. As
shown in Figure \ref{fig:overview}, we systematically outline the framework for using
Foundation Models (FMs) for recommender systems (FM4RecSys), covering everything
from the characteristics of recommendation data to specific downstream tasks.
We analyzed existing publications from various perspectives and introduced new insights.
Finally, we further delve into the latest unresolved questions and potential opportunities in this area.

\textbf{Criteria for Collecting Papers:} We collected over 150 papers related to
Foundation Models for Recommender Systems. Initially, we searched
top-tier conferences and journals such as ICLR, NeurIPS, WWW, WSDM, SIGIR, KDD, ACL,
EMNLP, NAACL, RecSys, CIKM, TOIS, TORS, and TKDE to identify recent work. The primary
keywords used in our search included large language models for recommender systems, generative recommendation,
large language models, multi-modal recommendation, and agents for recommendation.

\textbf{Contributions of This Survey:} The aim of this survey is to conduct a thorough
review of the advancements in Foundation Models for Recommender Systems (FM4RecSys).
It offers a comprehensive overview that enables readers to quickly grasp and engage with the field of foundation model-based recommendations. This survey establishes the groundwork to encourage innovation in RSs and explore the depth of this research domain. It is intended for researchers and practitioners interested in RSs, providing them with a guide for selecting FMs to address recommendation tasks. 
In summary, the key contributions
of this survey are threefold: (1) it offers a detailed review of foundational models for recommendation and introduces a classification scheme to organize and position current work; 
(2) It provides an overview and summary of the state of the
arts; and (3) it discusses the challenges, open issues, and identifies new trends and future directions in this research area to expand the horizons of FM4RecSys research.

The rest of this paper is structured as follows: 
\textbf{Section 2} explores the characteristics of RS data by contrasting the foundational aspects of traditional sources—such as user interactions, sequential behaviors, and network connections—with the emerging significance of multi-modal data.
\textbf{Section 3} introduces the recent advances in FMs,
highlighting the strengths and limitations of these models, from their scalability
and generalization capabilities to their ability in different tasks. \textbf{Section
4} discusses representation learning within FMs for RSs, analyzing the
techniques employed to derive representations that encapsulate the RS data. \textbf{Section
5} examines integration approaches for FM4RecSys, focusing on strategies that incorporate
foundation models into RS pipelines. \textbf{Section 6} details the specific tasks FM4RecSys models are
designed to address, including top-n ranking, sequential recommendation, and so on,
while also identifying task-specific challenges and current solutions. \textbf{Section
7} reviews empirical findings related to the opportunities and observed impacts
of FM4RecSys, including their potential for enhancing scalability, efficiency, and
reducing biases. In \textbf{Section 8}, we discuss open challenges and future
research directions in FM4RecSys, highlighting unresolved issues in model robustness,
explainability, and computational efficiency, as well as proposing directions
for advancing this emerging field. Finally, \textbf{Section 9} concludes the paper
by summarizing the contributions of foundation models to RSs.
\section{Traditional RS Data Characteristics}\label{sec:unidata}


This section explores the diverse types of data used in the RS field. 
Traditional RS rely on structured sources, such as user demographics, explicit feedback, and behavioral histories, augmented by sequential and network data that capture dynamic interaction patterns \cite{RicciRS15,ZhangYST19}. In contrast, recent advances involve integrating multi-modal data, including text, images, audio, and video, which expands the scope for understanding both user preferences and item attributes. Together, these diverse sources provide a robust foundation for addressing various recommendation tasks \cite{dl4rssurvey}.

\subsection{Traditional RS Data}

\textbf{User and item information:} In traditional recommendation systems, user and item information form the foundation for modeling preferences and generating recommendations. User information typically includes demographic attributes (e.g., age, gender, location), explicit preferences (e.g., ratings, likes, and reviews), and behavioral history (e.g., purchase records and browsing logs) \cite{RicciRS15}. On the other hand, item information consists of metadata such as category, brand, price, and textual descriptions, as well as multimodal features derived from images, videos, or audio \cite{RicciRS15, ZhangYST19}.
In addition, user-item interaction matrices play a crucial role in collaborative filtering-based methods. Implicit feedback, such as clicks, dwell time, and purchase actions, is particularly valuable as it reflects natural user engagement patterns without requiring explicit ratings. Side information, including knowledge graphs and attribute embeddings, further enriches user and item representations, improving recommendation quality, particularly in sparse interaction scenarios.

\textbf{Sequential information:} Sequential information plays a key role in RS, capturing the temporal sequence of user interactions with items \cite{WangHWCSO19}. This data can be particularly informative when user identifiers are available, allowing for the construction of personalized recommendation models that leverage individual historical behaviors. In cases where user identifiers are absent, session-based data can be employed to model interactions within a single session, making it suitable for scenarios involving anonymous users or cold-start problems. Additionally, location-based data, such as check-ins at points of interest (POI) \cite{IslamMDA22}, can be integrated to provide context-aware recommendations, which are particularly effective in mobile and location-based services.

\textbf{Network data:} Network data refers to the complex relationships between entities, such as users, items, social connections, and citations, which can be exploited by RS to enhance recommendation accuracy \cite{ZhangYST19}. Citation networks, which map the relationships between academic papers through citations, are commonly used in academic recommendation systems \cite{BeelGLB16}. Social networks, capturing the interactions and connections between users, enable RS to recommend friends, content, or groups based on social proximity and shared interests \cite{YangGLS14}.

\subsection{Multimodal RS Data}

\textbf{Textual data:} Textual data represents another common source of information in RS. Various forms of textual content, including hashtags, news articles, reviews, and so on, are used to derive rich contextual insights. Hashtags, often employed in social media, facilitate the clustering of similar content and are useful for recommending related items \cite{BalineniA23}. News articles and headlines can be analyzed to offer users recommendations that align with their reading interests \cite{KarimiJJ18}. User-generated reviews provide deep insights into user preferences and item attributes, which are crucial for content-based recommendation approaches, enhancing the personalization of recommendations \cite{Emrul}.

\textbf{Visual data:} Visual data is increasingly leveraged in RS, particularly in domains where the aesthetic qualities of items are significant. Visual features extracted from images, such as color, shape, and texture, are utilized to recommend items with similar visual characteristics. This approach is particularly valuable in domains like fashion, design, and art, where visual similarity plays a central role in user preferences \cite{HeM16}.

\textbf{Audio data:} Audio data, particularly in the form of music, is another key input for RS. Audio features, including genre, tempo, and mood, are analyzed to generate recommendations that align with a user's listening history and preferences. This data source is integral to music streaming services, where personalized playlists and track recommendations are based on complex audio feature analysis \cite{KneesS13}.

\textbf{Video data:} Video data provides a rich source of information for RS, especially in the context of multimedia content recommendation. Features derived from video, such as visual style, genre, and content type, are used to suggest similar videos or to enhance content discovery mechanisms \cite{GeP17}. This data is particularly relevant for video streaming platforms, where user engagement is heavily driven by the timely and relevant recommendation of content.

\textbf{Multimodal Fusion:} Drawing from prior multi-source and multimodal data in RSs, multimodal fusion is recognized as a crucial component for FM4RecSys. Several fusion paradigms have been explored in recent literature. \textit{Early fusion} combines modality-specific features at the input or representation level; for instance, embeddings from a product’s image and textual description may be concatenated and jointly processed \cite{liu2021nova}, allowing the model to capture inter-modal correlations from the outset. In contrast, \textit{late fusion} processes each modality independently and merges their outputs at a later stage, typically via averaging, weighted summation, or gating mechanisms \cite{wei2019mmgcn}, offering flexibility but potentially missing fine-grained interactions. Hybrid fusion balances these strategies by preserving modality-specific pathways while introducing interaction layers at intermediate or output stages. For example, NOVA \cite{liu2021nova} adopts a non-invasive hybrid strategy that maintains separate branches for collaborative and content features, with final fusion at the prediction layer to preserve ID-based signals. Attention-based fusion has become a dominant paradigm for adaptive integration, where attention mechanisms dynamically assign weights to modalities or their internal components (e.g., text tokens or image patches) based on contextual relevance. \textit{Co-attention} or cross-modal attention mechanisms are particularly effective for aligning modalities; for instance, CMBF \cite{chen2021cmbf} uses cross-attention to align image and text representations, capturing complementary semantics between product visuals and descriptions. These techniques are often integrated into transformer-based architectures. Building on this, \textit{cross-modal Transformers} further enhance multimodal fusion by leveraging self-attention and cross-attention layers to learn joint representations across modalities, and have been successfully applied to both sequential and static recommendation tasks \cite{chen2022mkgnn, jeong2024camrec}. Additionally, \textit{modality-aware gating} and \textit{graph-based fusion} offer further enhancements. Graph-based approaches model user-item interactions as graphs, where item nodes are enriched by multimodal content, allowing modality-specific message passing and collaborative signal propagation. Models such as MMGCN \cite{wei2019mmgcn} and MGAT \cite{tao2020mgat2} utilize attention or gated mechanisms to adaptively integrate modality signals, while gating networks like those used in MARIO \cite{kim2022mario} dynamically adjust the influence of each modality based on user preferences or contextual signals, enabling personalized and context-aware fusion.

\begin{tcolorbox}[
    title=Key Difference: Traditional and Multmodal RS data,
    colback=white,          
    colframe=black,        
    coltitle=white,        
    colbacktitle=black,     
    fonttitle=\bfseries,    
    boxrule=0.75pt,         
    arc=2pt,                
    left=6pt, right=6pt,    
    top=6pt, bottom=6pt     
  ]
\begin{itemize}[leftmargin=*]
    \item Traditional RS data primarily relies on fundamental user and item information such as demographics, explicit preferences, and behavioral histories, while multimodal RS data integrates diverse modalities like text, images, audio, and video to significantly broaden the information landscape.
    \item Traditional RS data is typically structured in clear formats like user-item interaction matrices and defined attributes, whereas multimodal RS data is often unstructured or semi-structured, necessitating specialized feature extraction and projection into a unified latent space.
    \item Multimodal RS data requires more complex fusion techniques than traditional RS data due to the necessity of integrating heterogeneous modalities.
  \end{itemize}
\end{tcolorbox}

\section{Recent Advances in Foundation Models}

In this section, we will give a brief introduction to recent advances in FMs, Multimodal Foundation Models, and Foundation Model Agents.


\subsection{Foundation Models}

A foundation model is any model that is trained on broad data (generally using self-supervision at scale) that can be adapted (e.g., fine-tuned) to a wide range of downstream tasks; current examples include BERT \cite{DevlinCLT19}, GPT-3 \cite{Ouyang0JAWMZASR22}, and CLIP \cite{RadfordKHRGASAM21}. However, the sheer scale and scope of foundation models from the last few years have stretched our imagination of what is possible; for example, GPT-3 has 175 billion parameters and can be adapted via natural language prompts to do a passable job on a wide range of tasks despite not being trained explicitly to do many of those tasks. Building upon the success of GPT-3, which is the first model to encompass over 100B parameters, several noteworthy models have been inspired, including GPT-J \cite{GPTJ}, BLOOM \cite{BLOOM}, OPT \cite{SusanZhang}, Chinchilla \cite{JordanHoffmann}, and LLaMA \cite{LLama}. These models follow the similar Transformer decoder structure as GPT-3 and are trained on various combinations of datasets.
Owing to their vast number of parameters, fine-tuning LLMs for specific tasks, such as RS, is often deemed impractical. Consequently, two prevailing methods for applying LLMs have been established: in-context learning (ICL) \cite{dong2022survey} and parameter-efficient fine-tuning \cite{Yuhui}. ICL is one of the emergent abilities of LLMs empowering them to comprehend and furnish answers based on the provided input context, rather than relying merely on their pre-training knowledge. This
method requires only the formulation of the task description and demonstrations in natural language, which are then fed as input to the LLM. Notably, parameter tuning is not required for ICL. Additionally, the efficacy of ICL can be further augmented through the adoption of the chain-of-thought prompting, involving multiple demonstrations (describe the chain of thought examples) to guide the model’s reasoning process. ICL is the most commonly used method for applying LLMs to information retrieval. Parameter-efficient fine-tuning aims to reduce the number of trainable parameters while maintaining satisfactory performance. LoRA \cite{Yuhui}, for example, has been widely applied to open-source LLMs (e.g., LLaMA and BLOOM) for this purpose. Recently, QLoRA \cite{DettmersPHZ23} has been proposed to further reduce memory usage by leveraging a frozen 4-bit quantized LLM for gradient computation. Despite the exploration of parameter-efficient finetuning for various NLP tasks, its implementation in RS tasks
remains relatively limited, representing a potential avenue for future research.

\subsection{Multi-modal Foundation Models}

Recently, substantial advancements in multimodal Foundation Models (MFMs) have emerged, largely augmenting standard FMs to accommodate multimodal input and output environments through economical training strategies \cite{ZhangY0L0C024}. 
MM-FMs leverage large language models (LLMs) as central components for semantic understanding and reasoning in multimodal tasks.
With marketing models such as GPT-4(Vision) \cite{GPT4} and Gemini \cite{Gemini,Gemini15} demonstrating exceptional multimodal understanding and generation capabilities, there is a surge of interest among researchers in Multimodal Foundation Models.
Initial studies primarily revolved around multimodal content comprehension and text generation. This included image-text understanding, as seen in groundbreaking projects such as BLIP-2 \cite{JunnanICML}, BLIP-3 \cite{xGen-MM}, LLaVA \cite{Haotian}, MiniGPT4 \cite{Kunchang}, and OpenFlamingo \cite{openflagmingo}. Initiatives like VideoChat \cite{VideoChat}, Video-ChatGPT \cite{VideochatGPT}, and LLaMA-VID \cite{LiWJ24} took this a step further with video-text understanding. The audio-text understanding capabilities of MM-LLMs were also largely explored in projects like Qwen-Audio \cite{QwenAudio} and Qwen-Audio2 \cite{QwenAudio2}.
Subsequent research broadened the abilities of MM-LLMs to include specific modality outputs. Image-to--text output tasks come into the picture with efforts such as Kosmos-G \cite{KosmosG}, Emu \cite{EMU}, and MiniGPT-5 \cite{MiniGPT5}, while projects like SpeechGPT \cite{SpeechGPT} and AudioPaLM \cite{AudioPalm} herald the advent of speech/audio-text output.
In recent times, the focus has been on imitating human-like any-to-any modality conversion, giving a glimpse into the potential path toward artificial general intelligence. Some attempts have been made to combine LLMs with external tools to accomplish comprehensive multimodal comprehension and generation, as showcased by VisualChatGPT \cite{VisualChatGPT}, HuggingGPT \cite{Yongliang}, and AudioGPT \cite{AudioGPT}.
In contrast, to minimize the cascading errors in the system, novel initiatives such as NExT-GPT \cite{NextGPT}, CoDi-2 \cite{CoDi-2}, and ModaVerse \cite{ModaVerse} have been developed. These provide end-to-end MM-LLMs and cover a full spectrum of modalities, signifying a promising step towards effectively modeling multimodal content in the RS field. 

\subsection{Foundation Model Agents}


The rapid evolution of LLM-based AI has spurred significant advancements in Agent AI, fundamentally reshaping how systems interact with complex environments. In recent years, researchers have equipped LLM agents with core components---memory, planning, reasoning, tool utilization, and action execution---that are essential for autonomous decision-making and dynamic interaction \cite{durante2024agent}. 

Single-agent systems leverage a unified model that integrates multiple interdependent modules.\footnote{
\url{https://github.com/huggingface/smolagents}}\footnote{ \url{https://www.langchain.com/langgraph}
}
The memory component acts as a structured repository that stores and retrieves contextually relevant information, such as user preferences and historical interactions \cite{zhang2024survey}. This persistent memory is crucial for maintaining coherent, long-term interactions and forms the foundation for personalization in recommendation settings.
The planning module is closely linked with advanced reasoning capabilities. Recent research has identified approaches such as task decomposition, multi-plan selection, external module-aided planning, reflection and refinement, and memory-augmented planning \cite{huang2024understanding}.
These techniques enable an agent to break down complex tasks, select and refine strategies based on evolving contexts, and leverage external knowledge sources.
Integrated reasoning further enhances decision-making by allowing the system to adapt dynamically to novel scenarios. Frameworks like the ReAct \cite{yao2022react} and Reflexion \cite{shinn2023reflexion} exemplify how interleaving reasoning with concrete actions, such as web-browsing or tool invocation, can significantly improve system robustness and adaptability.
Beyond internal cognitive processes, these agents increasingly rely on tool utilization to interface with external data and services. Systems like WebGPT \cite{hilton2021webgpt} illustrate the effectiveness of using external modules (e.g., web search engines) to retrieve real-time information. 
Other works, such as Retroformer \cite{yao2023retroformer} and AvaTaR \cite{wu2024avatar}, further optimize these interactions through policy gradient optimization and contrastive reasoning, respectively, to fine-tune tool usage and enhance performance over time. 

In contrast, LLM-based multi-agent systems emphasize collaboration among diverse autonomous agents. These systems are designed to mimic complex human workflows by facilitating inter-agent communication, task specialization, and coordinated decision-making.
Frameworks such as CAMEL \cite{li2023camel} and AutoGen demonstrate how agents with distinct roles can interact to solve problems more efficiently than a single, monolithic agent. By assigning specialized functions, ranging from ideation and planning to evaluation, these frameworks enable a division of labor that enhances overall system capability and flexibility.
Further advancements are seen in approaches like MetaGPT \cite{hong2023metagpt} and AgentLite \cite{liu2024agentlite}, which incorporate meta-programming techniques and lightweight libraries to dynamically allocate roles and coordinate complex workflows. These structured interactions not only improve task efficiency but also offer robustness in dynamic problem-solving environments.
Recent developments also include systems such as ChatEval \cite{chan2023chateval} and ChatDev \cite{qian2023communicative}, which leverage inter-agent debate and evaluative feedback to produce more nuanced and reliable outputs. This human-like discussion among agents is particularly beneficial in open-ended natural language generation tasks and complex software development processes.

\section{Representation Learning}\label{representation_learning}

In the pre-foundation model era, RS heavily relied on user and item representations from one-hot encoding for deep learning models. With the advent of FM4RecSys, there is a shift towards embracing more diverse inputs such as user profiles, item side information, and external knowledge bases like Wikipedia for enhanced recommendation performance. 
To be specific, numerous works\cite{BaoZZWF023,HuaLXCZ23} have identified that the key to building FM-based recommenders lies in bridging the gap between FMs' pre-training and recommendation tasks. To narrow the gap, existing work usually represents recommendation data in natural language for fine-tuning on FMs~\cite{Yaochen}. 
In this process, each user/item is represented by a unique identifier (e.g., user profile, item title, or numeric ID), and subsequently, the user’s historical interactions are converted into a sequence of identifiers. FMs can be fine-tuned on these identifiers to learn their representations to excel at recommendation tasks. 
Current recommendation data representation methods can be categorized as ID-based representation, multi-modal representation, graph representation, and hybrid representation.

\begin{table*}[ht]
\centering
\renewcommand{\arraystretch}{1.25}
\setlength{\tabcolsep}{4pt}
\small
\begin{adjustbox}{width=\textwidth}
\begin{tabular}{p{3.2cm}|p{4.5cm}|p{4.5cm}|p{4.5cm}|p{4.5cm}}
\toprule
\rowcolor{gray!20}
\textbf{Representation Learning Approach} & \textbf{Suitable for FM Paradigm} & \textbf{Suitable for Recommendation Tasks} & \textbf{Key Benefits} & \textbf{Challenges} \\
\midrule
\textbf{ID-Based Representation} &
\ding{228} Primarily Feature-Based &
\ding{228} Top-N Recommendation \newline
\ding{228} Basic Sequential Recommendation &
\ding{228} High efficiency and scalability \newline
\ding{228} Easy integration with existing pipelines &
\ding{228} Lack of semantic depth \newline
\ding{228} Limited adaptability in complex scenarios \\
\midrule
\textbf{Multi-modal Representation} &
\ding{228} Complementary to Feature-Based and Generative &
\ding{228} Cross-Domain Recommendation \newline
\ding{228} Item/Content Generation \newline
\ding{228} Conversational Recommendation &
\ding{228} Rich semantic context from diverse sources \newline
\ding{228} Enhanced handling of cold-start problems &
\ding{228} Increased preprocessing complexity \newline
\ding{228} Alignment issues and potential noise \\
\midrule
\textbf{Graph-Based Representation} &
\ding{228} Most suitable for Generative, Agentic Framework &
\ding{228} Sequential Recommendation \newline
\ding{228} Cross-Domain Recommendation \newline
\ding{228} Conversational Recommendation &
\ding{228} Captures rich relational and contextual information \newline
\ding{228} Facilitates dynamic adaptation via interactive modeling &
\ding{228} High computational cost \newline
\ding{228} Scalability issues in large-scale graphs \\
\midrule
\textbf{Hybrid Representation} &
\ding{228} Versatile across Feature-Based, Generative, and Agentic frameworks &
\ding{228} Applicable to a wide range of tasks including Top-N, Sequential, \newline
\ding{228} Conversational, Cross-Domain, and Item/Content Generation &
\ding{228} Balances efficiency with rich semantic expressiveness \newline
\ding{228} Leverages complementary strengths of multiple modalities &
\ding{228} Increased model complexity \newline
\ding{228} Challenges in effective fusion and alignment \\
\bottomrule
\end{tabular}
\end{adjustbox}
\caption{Comparative analysis of different representation learning approaches. The table shows the suitability of each approach for specific FM paradigms (Section~5) and recommendation tasks (Section~6), along with their key benefits and challenges.}
\label{tab:rep_learning_analysis}
\end{table*}

\subsection{ID-based Representation}\label{rec-id_representation}
In FM context, recent studies on ID-based representation utilize numeric IDs like ``[prefix]+[ID]'' (e.g., ``user\_123'' or ``item\_57'') to represent users and items, effectively capturing the uniqueness of items~\cite{geng2022recommendation,hua2023index}. Nevertheless, numeric IDs lack semantics and fail to leverage the rich knowledge in FMs. Furthermore, FMs require sufficient interactions to fine-tune each ID representation, limiting their generalization ability to large-scale, cold-start, and cross-domain recommendations. 
In addition, ID indexing necessitates updates to vocabularies to handle out-of-vocabulary (OOV) issues and parameter updates of the FMs that incur extra computational costs, highlighting the need for more informative representations.
Meanwhile, sequential ID indexing \cite{hua2023index} is utilized to capture the collaborative information in an intuitive way.

\subsection{Multi-modal Representation}\label{multi-modal_representation}
A promising alternative way lies in leveraging multi-modal side information, including utilizing images~\cite{SarkarBVLBLM23} (such as item visuals), textual content~\cite{recformer,ZhangW23,userembedding24,Joyce24,ShenGao24,UserLLM} 
(encompassing item titles, descriptions, and reviews), multi-modal elements~\cite{ShenZXJ22,YouwangKO22,HuangS0023,HuangW0023} 
(like short video clips and music), and external knowledge sources~\cite{ZhaiZWL023,Yunjia,KELLMRec} (such as item relationships detailed in Wikipedia). 
Yuan \textit{et al.}~\cite{YuanYSLFYPN23} underscores the advantages of multi-modality-based RS when compared to ID-based counterparts, drawing attention to the performance gains, which emphasizes that richer side information about users and items can enhance performance in cross-domain and cold-start recommendation scenarios.

However, the alignment between pure item side information and user-item interactions may not always be consistent \cite{Yaochen,Jiayi}. In other words, two items with similar visual or textual features might not necessarily share similar interaction patterns with users.
Thus, it is natural to utilize the hybrid representation that combines ID and multi-modal side information to achieve distinctiveness and semantic richness. For instance, TransRec~\cite{Xinyu} utilizes multi-faceted identifiers that combine IDs, titles, and attributes to achieve both uniqueness and semantic richness in item representation. CLLM4Rec~\cite{Yaochen} extends the vocabulary of FMs by incorporating user/item ID tokens and aligning them with user-item review text information through hard and soft prompting, allowing for accurate modeling of user/item collaborative information and content semantics. 

\subsection{Graph Representation}\label{graph_representation}
Moreover, the graphical structure inherent in user-item interactions makes graph representation a natural way to harness structured information from the user-item graph, thereby enhancing the recommendation system's ability to model complex interactions and user preferences~\cite{Wang0WFC19}.
LLM-based graph representation can primarily be categorized into two types. The first type involves encoding entity IDs (including user IDs, item IDs, and attribute IDs) using GNNs/HGNNs(Heterogeneous Graph Neural Networks) to obtain the graph representation. Then, LLMs are used to encode the textual descriptions of the entities themselves to obtain semantic text representations. These two representations are then fused to form a hybrid representation~\cite{Naicheng,Nurendra24,Qianzhao,Zhixuan24}. 
Ren \textit{et al.}\cite{RenWXSCWY024} incorporate auxiliary textual signals obtained through LLMs and align semantic spaces with collaborative relational graph signals.
The second type first utilizes LLMs to extract the text representation of the entities. Subsequently, based on the relationships between entities, GNNs/HGNNs are used to encode and obtain the graph representation.
Damianou \textit{et al.}\cite{Andreas24} train an HGNN on an item-item graph, where items are connected if they have been co-interacted with by the same user. Each node is associated with text embedding node features derived from LLMs applied to the item's description. The final representation combines the item's semantic information with the item-item relational representation. 
These integrations aim to harness the strengths of LLMs in natural language understanding and GNNs in relational data processing, resulting in a more powerful RS that can understand and recommend items.

In summary, Table \ref{tab:rep_learning_analysis} presents a comprehensive comparative analysis of various representation learning approaches as discussed in this section. The table categorizes ID-based, multi-modal, graph-based, and hybrid representations by aligning them with the three FM paradigms, Feature-Based, Generative, and Agentic, and further maps their applicability to specific recommendation tasks such as Top-N, Sequential, Conversational, Cross-Domain, and Item/Content Generation. This overview highlights that while ID-based representations are highly efficient and well-suited for fast, scalable applications like Top-N recommendation, multi-modal methods offer richer semantic context beneficial for cross-domain and content generation tasks. Similarly, graph-based approaches excel in capturing relational and dynamic user-item interactions, which are critical for sequential and conversational recommendations, while hybrid methods aim to combine these strengths to balance efficiency with expressive power. Overall, this table not only underscores the trade-offs associated with each representation strategy but also serves as a foundation for understanding their role within the integrated frameworks outlined in later sections.

\section{FM4RecSys: Integration Approaches}
\label{sec: integration approaches}



As mentioned in Section \ref{sub:paradigms}, we identified three paradigms of integration in current research: Feature-Based, Generative, and Agentic. 
Table \ref{tab:fm4recsys} provides a detailed comparison of these paradigms from different perspectives. Next, we delve deeper into each paradigm, surveying representative methods and discussing their strengths and limitations.

\begin{table*}[t]
\centering
\renewcommand{\arraystretch}{1.25}
\setlength{\tabcolsep}{4pt}
\small
\begin{adjustbox}{width=\textwidth}
\begin{tabular}{p{3.2cm}|p{4.5cm}|p{4.5cm}|p{4.5cm}|p{4.5cm}}
\toprule
\rowcolor{gray!20}
\textbf{FM Paradigm} & \textbf{Capabilities} & \textbf{Tasks} & \textbf{Key Benefits} & \textbf{Challenges} \\
\midrule
\textbf{Feature-Based RecSys} &
\ding{228} FM embeddings enhance user-item representation. \newline
\ding{228} Improves similarity matching \& cold-start handling. &
\ding{228} Top-N Recommendation \newline
\ding{228} Sequential Recommendation &
\ding{228} Better generalization to unseen items. \newline
\ding{228} Cold-start user/item adaptation. \newline
\ding{228} Improved user profiling. &
\ding{228} \textit{Limited Reasoning Capability} \newline
\ding{228} \textit{High inference cost of feature computation due to FM size}
\\
\midrule
\textbf{Generative RecSys} &
\ding{228} Generates personalized recommendations instead of ranking. \newline
\ding{228} Explains recommendations in natural language. &
\ding{228} Sequential Recommendation \newline
\ding{228} IN-context Recommendation \newline
\ding{228} Conversational Recommendation \newline
\ding{228} Explainable Recommendation &
\ding{228} Self-adaptive item generation. \newline
\ding{228} Personalized user experience. \newline
\ding{228} Multimodal RS enabled (text, image, audio). &
\ding{228} \textit{Bias Amplification in Generation} \newline
\ding{228} \textit{Explainability \& Trust Issues} \newline
\ding{228} \textit{High inference cost due to FM size} \newline
\\
\midrule
\textbf{Agentic RecSys} &
\ding{228} Acts as a decision-making agent. \newline
\ding{228} Remembers user history \& self-improves. \newline
\ding{228} Supports multi-turn interactions \& goal-driven planning. &
\ding{228} Long-term (Sequential) Personalization \newline
\ding{228} Multi-turn Adaptive Recommendation \newline
\ding{228} Planning \& Memory-Driven RS &
\ding{228} Context-aware and real-time learning. \newline
\ding{228} Continuous user adaptation \& memory. \newline
\ding{228} Proactive reasoning \& negotiation. &
\ding{228} \textit{Scalability \& Real-time Inference Cost} \newline
\ding{228} \textit{Evaluation Challenge of Agent Effectiveness}
\\
\bottomrule
\end{tabular}
\end{adjustbox}
\caption{Table~\ref{tab:fm4recsys} provides a detailed comparison of foundation model-powered recommender systems (FM4RecSys) across three major paradigms: Feature-Based, Generative, and Agentic. It summarizes their capabilities, tasks, key benefits, and challenges.}
\label{tab:fm4recsys}
\end{table*}

\subsection{Feature-Based Paradigm: Foundation Models as Feature Enhancers}
\label{subsec: feature-based paradigm}



\begin{figure}[h]
    \centering
    \includegraphics[width=1.0\linewidth]{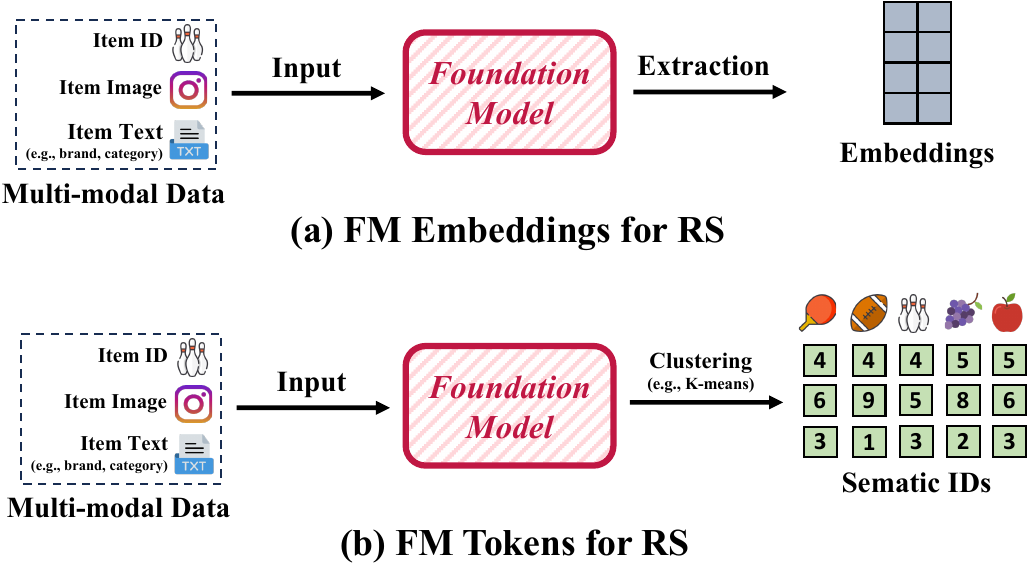}
    \caption{Examples of FM embeddings and tokens for RSs.}
    \label{fig:sec5_1}
\end{figure}

As shown in Figure \ref{fig:sec5_1}, existing works along this line can be categorized into two directions: (i) FM Embeddings for RSs, where FMs act as feature encoders to generate high-quality user/item embeddings for conventional recommendation models; and (ii) FM Tokens for RSs, where FMs generate semantic-aware tokens or indices to facilitate token-level generation and retrieval in recommendation.

\textbf{FM Embeddings for RS: } This modeling paradigm views the language model as a feature extractor, which feeds the features of items and users into LLMs and outputs corresponding embeddings. A traditional RS model can utilize knowledge-aware embeddings for various recommendation tasks. 
RLMRec \cite{RenWXSCWY024} treats LLMs as a text encoder to map items or users into a semantic space and aligning their semantic space with collaborative relation modeling for better representation learning. They leverage the advanced text comprehension abilities of LLMs to capture the nuanced semantic aspects of user behaviors and preferences. AlphaRec \cite{AlphaRec} employs linear mapping to project language representations of item titles into a behavior space for recommendation. Those language representations are transformed into high-quality behavior representations, resulting in outstanding recommendation performance. LLMRec \cite{wei2024llmrec} uses LLMs for graph augmentation for RecSys by augmenting user-item interaction edges, item node attributes, and user node profiles. This work addresses the scarcity of implicit feedback signals by enabling LLMs to explicitly reason about user-item interaction patterns. BinLLM \cite{zhang2024text} encodes collaborative information textually for LLM-based recommendation by converting collaborative embeddings into binary sequences. Specifically, BinLLM transforms collaborative embeddings from external models into binary sequences, a text format compatible with LLMs, enabling the direct utilization of collaborative information in a text-like format. iDreamRec \cite{hu2024generate} makes further steps to combine LLMs embeddings with Diffusion Models \cite{ho2020denoising} for recommendation tasks by incorporating text embeddings from LLMs to accurately model item distributions. Specifically, given the metadata about items (e.g., titles), this work prompts GPT to generate detailed textual descriptions, offering richer content than simple item IDs.


\textbf{FM Tokens for RS:} 
This modeling paradigm generates tokens based on the input items' and users' features. The generated tokens capture potential preferences through semantic mining, which can be integrated into the decision-making process of a recommendation system. Recent studies have introduced methods in which items are represented by semantically meaningful tokens, enabling more accurate and context-aware recommendations. 
For instance, the TIGER framework \cite{rajput2023recommender} proposes a generative retrieval approach that autoregressively decodes item identifiers. In this framework, each item is assigned a \emph{Semantic ID}---a tuple of codewords derived from the item's content features---thus allowing the system to predict the next item a user might interact with based on previous interactions. Similarly, the LC-Rec model \cite{zheng2024adapting} addresses the semantic gap between large language models and recommender systems by integrating both language and collaborative semantics. It employs a learning-based vector quantization method to assign meaningful item indices, enabling the language model to generate items directly from the entire item set without relying on predefined candidates. Other frameworks extend these ideas further. For example, ColaRec \cite{wang2024content} combines content information and collaborative signals within a unified sequence-to-sequence generative framework, while the EAGER framework \cite{wang2024eager} integrates behavioral and semantic information through a two-stream generative architecture. Moreover, the COBRA framework \cite{yang2025sparse} adopts a cascaded approach that alternates between sparse semantic IDs and dense vectors to capture both semantic insights and collaborative signals from user-item interactions. OneRec \cite{deng2025onerec} explores related techniques to further enhance recommender systems by leveraging token generation via large language models, thereby highlighting the potential of semantic tokenization in this domain.

\subsection{Generative Paradigm: Foundation Models as Generators of Recommendations}
\label{subsec: generative paradigm}





\begin{figure}[ht]
    \centering
    \includegraphics[width=1\linewidth]{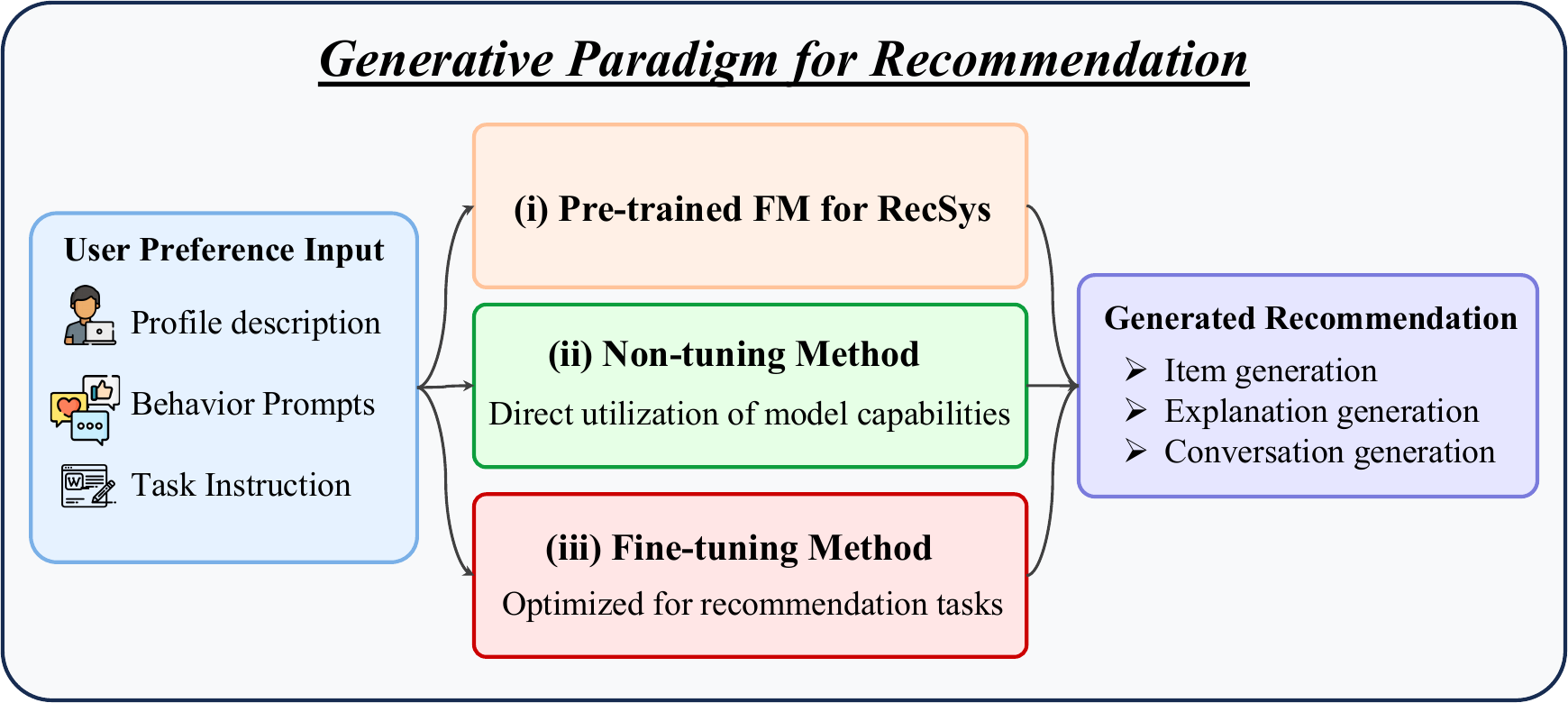}
    \caption{An illustration of the generative paradigm for recommendation. 
    User preference inputs (e.g., the profile description, behavior prompts, and task instructions) are utilized to guide the pre-trained foundation models (FM) for RS. 
    The model can be leveraged in a non-tuning manner by directly utilizing its capabilities or via fine-tuning for specific recommendation tasks, producing various forms of generated recommendations such as item generation, explanation generation, and conversation generation.}
    \label{fig:framework-generative-recsys}
\end{figure}

Different from the feature-based paradigm, the generative paradigm formulates recommendations as a generative task and provides diverse recommendations to meet user preferences. Specifically, it transforms pre-trained generative models into powerful end-to-end recommendation systems. The input of FMs typically includes varying user preference information, such as profile description, behavior prompt, and task instruction, while the output is expected to generate reasonable recommendations. In this section, we mainly focus on how to transfer an FM into a recommender, and thereby classify related research works into three categories: 
(1) Pre-trained FM for RS, (2) Non-tuning FM4RecSys, and (3) Fine-tuning FM4RecSys as shown in Figure \ref{fig:framework-generative-recsys}.

\textbf{Pre-trained FM4RecSys:} Few works like M6 Rec \cite{cui2022m6}, PTUM \cite{wu2020ptum} and RecGPT \cite{ngo2024recgpt}  pre-train the whole model on massive recommendation datasets by adopting transformer-based models for next-item prediction and applying different language modeling tasks, such as masked language modeling, permutation language modeling, and so on. This line of work generally requires a large amount of domain data for RS, leading to high training costs. 

\textbf{Non-tuning FM4RecSys:}  Foundation models (FMs) have demonstrated strong zero- and few-shot capabilities across many tasks \cite{li2023your,brown2020language,Ouyang0JAWMZASR22}. Consequently, recent studies assume that FMs inherently possess recommendation capabilities and aim to activate these abilities through the use of tailored prompts where FMs parameters remain unchanged. Due to the rich textual side information, prompting with LLMs has demonstrated powerful recommendation ability. Non-tuning FM4RecSys focuses on designing appropriate prompts to stimulate the recommendation abilities of LLMs. Liu \textit{et al.}~\cite{benchmarking} propose a prompt construction framework to evaluate the ability of ChatGPT on five common recommendation tasks, providing zero-shot and few-shot versions for each type of prompt. He \textit{et al.} \cite{hou2024zero} not only use prompts to evaluate the ability of LLMs on sequential recommendation but also introduce recency-focused prompting and in-context learning strategies to alleviate order perception and position bias issues of LLMs. More recently, some works~\cite{prompting_survey} have focused on designing novel structures in prompting for FM4RecSys. Yao \textit{et al.}~\cite{knowledgeplugin} include heuristic prompts including item attributes in natural language, along with collaborative filtering information presented through text templates and knowledge graph reasoning paths. Similarly, Rahdari \textit{et al.} \cite{Behnam23} crafte hierarchical prompt structures that encapsulate information about recommended items and top-k similar item information in the user interaction history.
Hou \textit{et al.} \cite{hou2024zero} leverage the zero-shot and in-context learning capabilities of LLMs to construct prompts for sequential movie recommendations. However, using LLMs as recommender systems without pre-training or fine-tuning, only depending on in-context learning, still lags behind traditional supervised methods like SASRec \cite{SARS} in sequential tasks \cite{hou2024zero}. The primary reason is that LLMs have limited ability to perceive the order of historical interaction sequences. As the length of these sequences increases, the performance of LLM recommendations tends to decline. Therefore, optimizing the long-sequence modeling capability of LLMs presents a potential opportunity to enhance their effectiveness in recommendation scenarios.

\textbf{Fine-tuning FM4RecSys:} 
This line of work generally requires a large amount of domain-specific data for retraining the FMs, leading to extra computational costs. Although FMs generally exhibit strong zero- and few-shot capabilities, it is unsurprising that they may fall short of outperforming recommendation models specifically trained on task-related data for a given task. Therefore, one straightforward approach is to fine-tune powerful FMs, which contain world knowledge, using task-specific data for downstream recommendation tasks. 
Recently, there has been extensive exploration of fine-tuning large language models (LLMs) for recommendation tasks.
TCF~\cite{RuyuLi2023} adopts and fine-tunes LLMs to create a universal item representation for recommendation tasks, contrasting this approach with the increasingly popular prompt-based approach using ChatGPT. Unfortunately, despite utilizing an item encoder with tens of billions of parameters, it still requires re-adaptation for new data to achieve optimal recommendations. Furthermore, this type of model has not demonstrated the strong transferability that was expected, suggesting that constructing large-scale foundational recommender models may be more challenging than in the fields of CV and NLP.
InstructRec~\cite{InstruRec} designs abundant instructions for tuning, including 39 manually designed templates with preference, intention, task form, and context of a user. 
After instruction tuning, LLMs can understand and follow different instructions for recommendations.
TallRec~\cite{TallRec} uses LoRA~\cite{lora}, a parameter-efficient tuning method, to handle the two-stage tuning for LLMs. It is first fine-tuned on the general data of Alpaca \cite{alpaca}, and then further fine-tuned with the historical information of users. It utilizes item titles as input and shows effectiveness for cold-start recommendations.
BIGRec~\cite{BigRec} emphasizes that LLMs struggle to integrate statistical data such as popularity and collaborative filtering because of their inherent semantic biases. To address this, BIGRec fine-tunes LLMs through instruction tuning to produce tokens that symbolize items. However, aligning LLM outputs with real-world items is challenging due to their inventive nature. BIGRec subsequently aligns these generated tokens with real items in the recommendation database by incorporating statistical data like item popularity.DEALRec~\cite{efficientFT} has introduced a data pruning method that employs two scores: the influence score, which estimates the impact of sample removal on performance using a small surrogate model, and the effort score, which prioritizes challenging samples for LLMs. This approach enables efficient fine-tuning of LLMs, thereby enhancing both efficiency and accuracy.

Meanwhile, Diffusion Models (DMs) \cite{ho2020denoising,rombach2022high} are another generative FMs that have recently started being applied to customize the best visual recommendation content to different users. It is non-trivial to consider both the user’s potential preferences, based on historical and contextual information, and the visual coherence and correlation of content. The emergence of visual FMs, particularly Stable Diffusion \cite{rombach2022high}, offers a promising direction for automating and even personalizing item display content generation. DiFashion \cite{xu2024diffusion} can not only generate complementary fashion items but also create personalized outfit images from scratch based on user preferences. The model fine-tunes the latest Stable Diffusion model to ensure high fidelity, compatibility, and personalization in the generated fashion images. AdBooster \cite{shilova2023adbooster} introduces the generative creative optimization task and leverages user interest signals to personalize advertisement creative generation using the outpainting technique of Stable Diffusion. The subsequent work, CG4CTR \cite{yang2024new}, further refine the solution by introducing a new automated creative generation for the click-through rate (CTR) optimization pipeline. Specifically, it employs the inpainting mode of Stable Diffusion to generate background images while preserving the main product details. More recently, DynaPIG \cite{czapp2024dynamic} leverages diffusion models to generate visually appealing personalized product images, enhancing user engagement with recommendations. However, since diffusion-based FMs are primarily designed for vision tasks, they are not inherently suited for item recommendation. Many recent diffusion-based works \cite{wang2023diffusion,yang2023generate,li2023diffurec,huang2024dual} tend to train generative recommender from scratch for domain-specific tasks. Despite these advancements, a significant gap remains in effectively integrating the strengths of diffusion-based FMs with broader frameworks for recommendation. To bridge this gap, researchers have begun exploring how FMs can act not only as generative models but also as central components in autonomous systems, enabling more dynamic and interactive recommendation experiences.






\subsection{Agentic Paradigm: Foundation Models as Interactive Recommender Agents}
\label{subsec: agentic paradigm}



\begin{figure*}[ht]  
	\centering
        \includegraphics[width=1\linewidth]{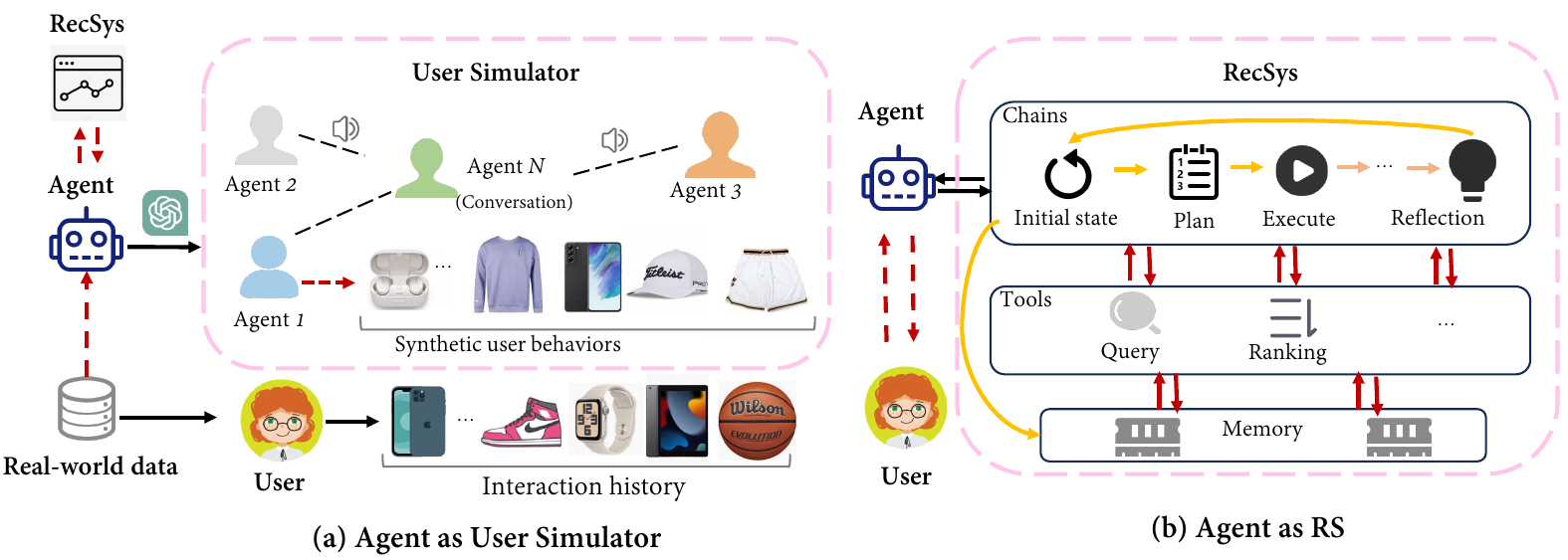}
	\caption{Two types of personalized agents in FM4RecSys: (a) Agent as User Simulator and (b) Agent as Recommender System. }
	\label{fig:agent}
\end{figure*}

In the FM-powered autonomous agent system, FMs function as the agent’s brain, complemented by key components like planning, memory, and tool use~\cite{weng2023prompt}. 
There are many inspiring works, such as AutoGPT and BabyAGI proving the FM-based agents' potential. These agents can store their past experiences and make better decisions for future behaviors.
In RS scenarios, agents are typically represented as either User Simulators or the Recommender System itself, as shown in Figure~\ref{fig:agent}.

\textbf{{Agent as User Simulator}:} 
This paradigm uses agents to simulate user behaviors for real-world recommendations.
Gathering sufficient and high-quality user behavior data is expensive and ethically complex. Furthermore, real-world user interaction data is often very sparse, such as in the case of cold-start users.
Besides, traditional methods \cite{ZhuLLWGLC17,RecSim} 
often face challenges in simulating complex user behaviors due to models' capabilities, whereas FMs have demonstrated potential in this area \cite{wang2023recagent}.
Consequently, employing personalized agents powered by FMs for RSs emerges as a logical and effective strategy.
Wang \textit{et al.} \cite{wang2023recagent} treat each user as an FM-based autonomous agent within a virtual simulator named RecAgent. This simulator allows for the free interaction, behavior, and evolution of different agents, taking into account not only actions within the RS, like item browsing and clicking, but also external factors like social interactions. Zhang \textit{et al.} \cite{agent4rec} further investigate the extent to which FM-empowered generative agents can accurately simulate real human behavior for movie recommendation. They design Agent4Rec, a recommender system simulator with 1,000 LLM-empowered generative agents interacting with personalized movie recommendations in a page-by-page manner with various actions. 
Since previous FM methods perform poorly in long-term recommendations, Shi \textit{et al.}~\cite{BiLLP24} propose a Bi-level Learnable LLM Planner (BiLLP) framework to enhance these systems. BiLLP leverages LLMs to balance short-term and long-term user satisfaction by combining macro-learning for high-level planning and micro-learning for personalized actions. This framework shows significant potential in addressing the challenges associated with long-term recommendation planning.
After that, \cite{AgentCF} treats both users and items as agents and enables a collaborative learning process that optimizes the interactions between agents.
Recently, Zhang \textit{et al.} ~\cite{Erhan24} propose the USimAgent that can simulate users' querying, clicking, and stopping behaviors during search sessions through LLMs, and thus, is capable of generating complete search sessions for specific search tasks. BASES~\cite{BASES24} utilizes LLM-based agents for large-scale user simulation in web search, generating diverse user profiles and search behaviors. It has demonstrated effectiveness through evaluations on both Chinese and English benchmarks.
Meanwhile, Huang \textit{et al.} ~\cite{Feiran24} propose the LLM Interaction Simulator (LLM-InS), which simulates user behavior patterns based on content features. This approach transforms cold items into warm items, addressing the challenge of recommending cold items due to the lack of historical user interactions. Specifically, the LLM-InS simulator essentially functions as a CTR (Click-Through Rate) model. It takes information about a specific user and a specific item as input and predicts whether the user will click on the item. For a cold-start item, a subset of users is "recalled," and the simulator predicts whether these users will click on the cold-start item, thereby generating interaction data. This simulated interaction data is then used to update the item embeddings.

\textbf{Agent as RecSys:} This paradigm leverages the robust capability of FMs, including reasoning, reflection, and tool usage for recommendation.
RAH \cite{Yubo23} improves alignment with user personalities and mitigate biases by incorporating FM-based agents and a Learn-Act-Critic loop. Then, Wang \textit{et al.}~\cite{wang2023recmind} first introduce a Self-Inspiring planning algorithm that keeps track of all past steps of the agent to help generate new states. At each step, the agent looks back at all the paths it has taken before to figure out what to do next. This approach aids in employing databases, search engines, and summarization tools, combined with user data, for producing tailored recommendations.
InteRecAgent \cite{InteRecAgent} models the FMs as the brain and recommendation models as tools providing domain-specific knowledge, enabling FMs to parse user intent and generate responses.
They specify a core set of tools essential for RS tasks, Information Query, Item Retrieval, and Item Ranking, and introduce a candidate memory bus, allowing previous tools to access and modify the pool of item candidates.

\textbf{Agent both for Simulator and RecSys:}
More recently, to address the gap in multi-agent collaboration within recommendation systems, Wang \textit{et al.}~\cite{Zhefan24} introduce MACRec, a framework that enhances recommendation tasks through the collaboration of specialized agents like Manager, User/Item Analyst, Reflector, Searcher, and Task Interpreter. MACRec can be applied to various tasks, including rating prediction, sequential recommendation, conversational recommendation, and explanation generation, tackling recommendation tasks through the collaboration of various agents. 
Cai \textit{et al.}~\cite{cai2024large} propose the PUMA framework, which uses a memory system to retrieve relevant past user interactions, enhancing the agent’s ability to align actions with user preferences.

\begin{table*}[t!]
\centering
\begin{tabular}{cccccc}
\toprule
\rowcolor{gray!20}
\textbf{Model} & \textbf{Objectives} & \textbf{Single-type Agents} & \textbf{Multi-type Agents}  & \textbf{Diverse Rec. Scenarios} & \textbf{Open-source} \\
\midrule
RecAgent~\cite{wang2023recagent} & User Simulation & \Checkmark &  &  & \Checkmark \\
Agent4Rec~\cite{agent4rec} & User  Simulation & \Checkmark &  &  &  \Checkmark \\
LLM-Ins~\cite{Feiran24} & User  Simulation & \Checkmark &  &  &   \\
PMG~\cite{ShenZ0ZX24} & User  Simulation & \Checkmark &  &  &  \Checkmark \\
BiLLP~\cite{BiLLP24} & User  Simulation & \Checkmark &  &  &  \Checkmark \\
BASES~\cite{BASES24} & User  Simulation & \Checkmark &  &  &   \\
USimAgent~\cite{Erhan24} & User  Simulation & \Checkmark &  &  &   \\
AgentCF~\cite{AgentCF} & U-I Inter Simulation & & \Checkmark &  & \Checkmark  \\
WebAgent~\cite{cai2024large} & U-I Inter Simulation & & \Checkmark & \Checkmark &  \Checkmark\\
\midrule
RAH~\cite{Yubo23} & Recommender &  & \Checkmark &  &  \\
RecMind~\cite{wang2023recmind}  & Recommender & \Checkmark &  & \Checkmark &  \\
InteRecAgent~\cite{InteRecAgent}  & Recommender & \Checkmark &  &  & \\
MACRec~\cite{Zhefan24} & Recommender  & \Checkmark & \Checkmark & \Checkmark & \Checkmark \\
\bottomrule
\end{tabular}
\vspace{0.5cm} 
\caption{Comparison among Foundation Model Agents for RecSys. Note that \textsl{Single-type Agents} indicate all agents serve the same role (e.g., users), while \textsl{Multi-type Agents} refer to agents having multiple roles and capabilities (e.g., managers, reflectors).}
\label{tab:agent_paper}
\end{table*}

To summarize, as shown in Table \ref{tab:agent_paper}, the simulation-oriented work focuses on using agents to simulate user behaviors and item characteristics in RSs. This line of research seeks to enhance the understanding of user preferences but lacks integration into RSs.
The goal of recommender-oriented studies is to build a ``recommender agent" with planning and memory components to tackle recommendation tasks.

\subsection{Critical Analysis}

\textbf{Feature-based RecSys:} In the feature-based framework, foundation models are mainly used as high-quality feature extractors to generate improved embeddings for users and items. This approach benefits from leveraging vast pre-trained knowledge, which enhances representation quality and facilitates better ranking in tasks like Top-N recommendation and sequential recommendation. However, a critical drawback is that these models function in an auxiliary capacity and remain largely decoupled from the core recommendation decision process. This separation limits the system’s ability to dynamically adapt to contextual shifts or user-specific feedback. Additionally, the computational overhead associated with employing large-scale foundation models for embedding extraction can be a concern in real-world scenarios. Although the modularity and ease of integration make the feature-based approach attractive for enhancing existing systems, its limited reasoning and interactive capabilities constrain its application in more complex, dynamic environments.

\textbf{Generative RecSys:} The generative framework redefines recommendation by transforming it into an end-to-end natural language generation problem. This approach takes advantage of foundation models’ inherent ability to generate personalized recommendations, natural language explanations, and even novel items, which proves beneficial in zero- or few-shot settings. Despite its promise, the generative paradigm faces significant challenges in terms of output controllability and alignment with user intent. The models’ emphasis on fluency can sometimes lead to recommendations that are less precise or relevant (e.g., the OOV problem: generated items are out of vocabulary), and the qualitative nature of generated content makes it difficult to evaluate performance using standard ranking metrics. Moreover, the training and inference processes in generative methods tend to be resource-intensive, thereby raising concerns about latency and scalability in practical applications. Thus, while the generative approach offers innovative avenues for personalization and explanation, its challenges in output quality control and computational cost must be addressed to ensure its viability in real-world recommendation systems.

\textbf{Agentic RecSys:} Agentic frameworks treat the recommender system as an autonomous agent capable of interactive decision making, memory retention, and real-time planning. This paradigm promises significant advancements by engaging in multi-turn dialogues, incorporating feedback, and even utilizing external tools to refine recommendations. However, its complexity poses substantial challenges. The integration of memory and planning components, while theoretically enabling dynamic adaptation, introduces issues regarding scalability and real-time performance. The unpredictable nature of autonomous decision-making can lead to inconsistencies in recommendation quality, and managing long-term user satisfaction while maintaining immediate responsiveness is a non-trivial problem. Additionally, the higher computational burden and the difficulties in designing robust evaluation methods for interactive systems are notable obstacles. Despite these limitations, the agentic approach is particularly promising for creating personalized experiences that more closely mimic human-like reasoning and decision-making.

\begin{tcolorbox}[
    title=Comparative Discussion,
    colback=white,          
    colframe=black,        
    coltitle=white,        
    colbacktitle=black,     
    fonttitle=\bfseries,    
    boxrule=0.75pt,         
    arc=2pt,                
    left=6pt, right=6pt,    
    top=6pt, bottom=6pt     
  ]
\begin{itemize}[leftmargin=*]
    \item \textbf{Feature-Based Framework:} Excels in simplicity and producing high-quality semantic embeddings, but lacks dynamic adaptability and interactive reasoning.
    \item \textbf{Generative Framework:} Provides personalized and context-rich outputs, yet struggles with output controllability and entails high computational costs.
    \item \textbf{Agentic Framework:} Offers advanced interactive capabilities and real-time adaptation, though it is hindered by complexity and scalability concerns.
\end{itemize}
\end{tcolorbox}

\section{FM4RecSys – Tasks}
\label{sec: tasks}

In this section, we first revisit the general formulation of the main recommendation tasks, including Top-N recommendation, sequential recommendation, conversational recommendation, and cross-domain recommendation, followed by the recent progress of FM4RecSys.

\begin{figure*}[t!]
\centering
\large
\begin{forest}
for tree={   
font=\small, 
draw=myblue, semithick, rounded corners,
       minimum height = 1.5ex,
        minimum width = 2em,
    anchor = west,
     grow = east,
forked edge,        
    s sep = 0.8mm,    
    l sep = 3.5mm,    
 fork sep = 2mm,    
           }
[The Classification of FM4RecSys, rotate=90, anchor=center
    [Taxonomy of Finding (\S~\ref{sec: explored opportunities and findings}), text width=3cm
        [In-context Learning in FM4RecSys (\S~\ref{subsec: in-context learning capability}), text width=2.5cm
            [{GPT-3, GPT-4~\cite{hou2024zero}, LLaMA, LLMRec~\cite{liu2023llmrec}, Flan-T5~\cite{zhang2023instruction}, LLM Reasoning Graphs~\cite{wang2023llmrg}, ChatGPT~\cite{spurlock2024chatgpt}}, text width=9.3cm, fill=mygrey]
        ]
        [Multi-modality in FM4RecSys (\S~\ref{subsec: multi-modal recommender systems}), text width=2.5cm
            [{M6Rec~\cite{cui2022m6rec}, Flamingo~\cite{alayrac2022flamingo}, Kosmos-2~\cite{peng2023kosmos}, VIP5~\cite{geng2023vip5}, MMRec~\cite{zhou2023mmrec}, MMREC~\cite{tian2024mmrec}, Video-LLaMA~\cite{zhu2023videollama}, PMG~\cite{shen2024pmg}, TalkPlay~\cite{he2024talkplay}, LLark~\cite{brookes2023llark}, CM3leon~\cite{beaumont2023cm3leon}}, text width=9.3cm, fill=mygrey]
        ]
        [Fairness for FM4RecSys (\S~\ref{subsec: fairness_for_FM4RecSys}), text width=2.5cm
            [{ User Group-side:\cite{abs-2305-12090,ZhangBZWF023,Yashar24}, Item-side:\cite{hou2024zero}\cite{JiangBZW0F024}}, text width=9.3cm, fill=mygrey]
        ]
        [Interpretability in FM4RecSys (\S~\ref{subsec: explanations and interpretation}), text width=2.5cm
            [{  NETE \cite{LiZC20}, M6Rec\cite{cui2022m6}, LLM4Vis \cite{LLM4Vis}, PosthocE \cite{Ngoc23}, LLMHG \cite{Zhixuan24}, ChatGPT4Rec \cite{liu2023first,liu2023chatgpt} }, text width=9.3cm, fill=mygrey]
        ]
    ]
    [Taxonomy of Task (\S~\ref{sec: tasks}), text width=3cm
        [Cross-domain in FM4RecSys (\S~\ref{sec:cross-domain}), text width=2.5cm
            [{  RPM \cite{HuSLC23}, HAMUR \cite{Xiaopeng}, LLM-Rec \cite{Zuoli}, S\&R Multi-Domain Foundation \cite{GongDSSLZ23}, Uni-CTR~\cite{Zichuan}, AdaT \cite{Junchen}, PFCR \cite{LeiGuo24}}, text width=9.3cm, fill=mycyan]
        ]
        [FM for Conversational RecSys (\S~\ref{sec:interactive}), text width=2.5cm
            [{ AVPPL \cite{Gangyi}, RecInDial \cite{lingzhi2021}, LLM4ZeroshotCRS \cite{HeXJSLFMKM23}, Spark \cite{spark}, RecMind \cite{wang2023recmind}, KyleCRS \cite{KyleCRS}, ChatGPT4Rec \cite{liu2023chatgpt}, PECRS \cite{RavautZXSL24}, Benchmark \cite{wang2023rethinking}}, text width=9.3cm, fill=mycyan]
        ]
        [FM for Sequential recommendation (\S~\ref{sec:context}), text width=2.5cm
            [{LLMSeqSim \cite{HarteZLKJF23}, KAR \cite{Yunjia}, DRDT \cite{Wangyureasoning}, PPR \cite{YiqingPPR}, KP4SR \cite{ZhaiZWL023}, LLaRA \cite{Jiayi}, LLM4SBR \cite{Shutong24}, CALRec \cite{Yaoyiran}}, text width=9.3cm, fill=mycyan]
        ]
        [Top-N Recommendation (\S~\ref{subsec: top-N recommendation}), text width=2.5cm
            [{IID \cite{hua2023index}, P5 \cite{geng2022recommendation}, OpenP5 \cite{Xu2023}, ChatGPT4Rec \cite{LiZM23,DaiSZYSXS0X23}, TASTE \cite{LiuMXLYL0023}, Llama4Rec \cite{Sichun24}, }, text width=9.3cm, fill=mycyan]
        ]
    ]
     [Taxonomy of Framework (\S~\ref{sec: integration approaches}), fit=band, text width=3cm
        [Agentic Paradigm (\S~\ref{subsec: agentic paradigm}), text width=2.5cm
            [Agent as User Simulator, text width=2.1cm
                [{RecAgent \cite{wang2023recagent}, Agent4Rec \cite{agent4rec}, BiLLP \cite{BiLLP24}, PMG \cite{ShenZ0ZX24}, AgentCF \cite{AgentCF}, USimAgent \cite{Erhan24}, BASES \cite{BASES24}, LLM-InS \cite{Feiran24}}, text width=6.5cm, fill=myyellow]
            ]
            [Agent as RS, text width=2.1cm
                [{RAH \cite{Yubo23}, RecMind \cite{wang2023recmind}, InteRecAgent \cite{InteRecAgent}, MACRec \cite{Zhefan24}}, text width=6.5cm, fill=myyellow]
            ]
        ]
        [Generative Paradigm (\S~\ref{subsec: generative paradigm}), text width=2.5cm
            [Pre-trained FM4RS, text width=2.1cm
                [{M6 Rec~\cite{cui2022m6}, PTUM~\cite{wu2020ptum}, RecGPT \cite{ngo2024recgpt}}, text width=6.5cm, fill=myyellow]
            ]
            [Non-tuning FM4RS, text width=2.1cm
                [{LLMRec \cite{benchmarking}, LLMRank \cite{hou2024zero}, Knowledge Plugins \cite{knowledgeplugin}, Logic-Scaffolding \cite{Behnam23},}, text width=6.5cm, fill=myyellow]
            ]
            [Fine-tuning FM4RS, text width=2.1cm
                [{TCF~\cite{RuyuLi2023}, InstructRec~\cite{InstruRec}, TallRec~\cite{TallRec},BIGRec~\cite{BigRec}, DEALRec~\cite{efficientFT}, DiFashion \cite{xu2024diffusion}, AdBooster \cite{shilova2023adbooster}, CG4CTR \cite{yang2024new}, DynaPIG \cite{czapp2024dynamic}}, text width=6.5cm, fill=myyellow]
            ]
        ]
        [Feature-Based Paradigm (\S~\ref{subsec: feature-based paradigm}), text width=2.5cm
            [FM Embeddings for RS, text width=2.1cm
                [{RLMRec \cite{RenWXSCWY024}, AlphaRec \cite{AlphaRec}, LLMRec \cite{wei2024llmrec}, BinLLM \cite{zhang2024text}, iDreamRec \cite{hu2024generate}, }, text width=6.5cm, fill=myyellow]
            ]
            [FM Tokens for RS, text width=2.1cm
                [{TIGER \cite{rajput2023recommender}, LC-Rec \cite{zheng2024adapting}, ColaRec \cite{wang2024content}, EAGER \cite{wang2024eager}, COBRA \cite{yang2025sparse}, OneRec \cite{deng2025onerec}}, text width=6.5cm, fill=myyellow]
            ]
        ]
    ]
]
\end{forest}
\caption{The taxonomy of Foundation Model (FM) for Recommendation Systems (FM4RecSys). Representative works are shown under each sub-category for reference. }
\label{fig:task_overview}
\end{figure*}
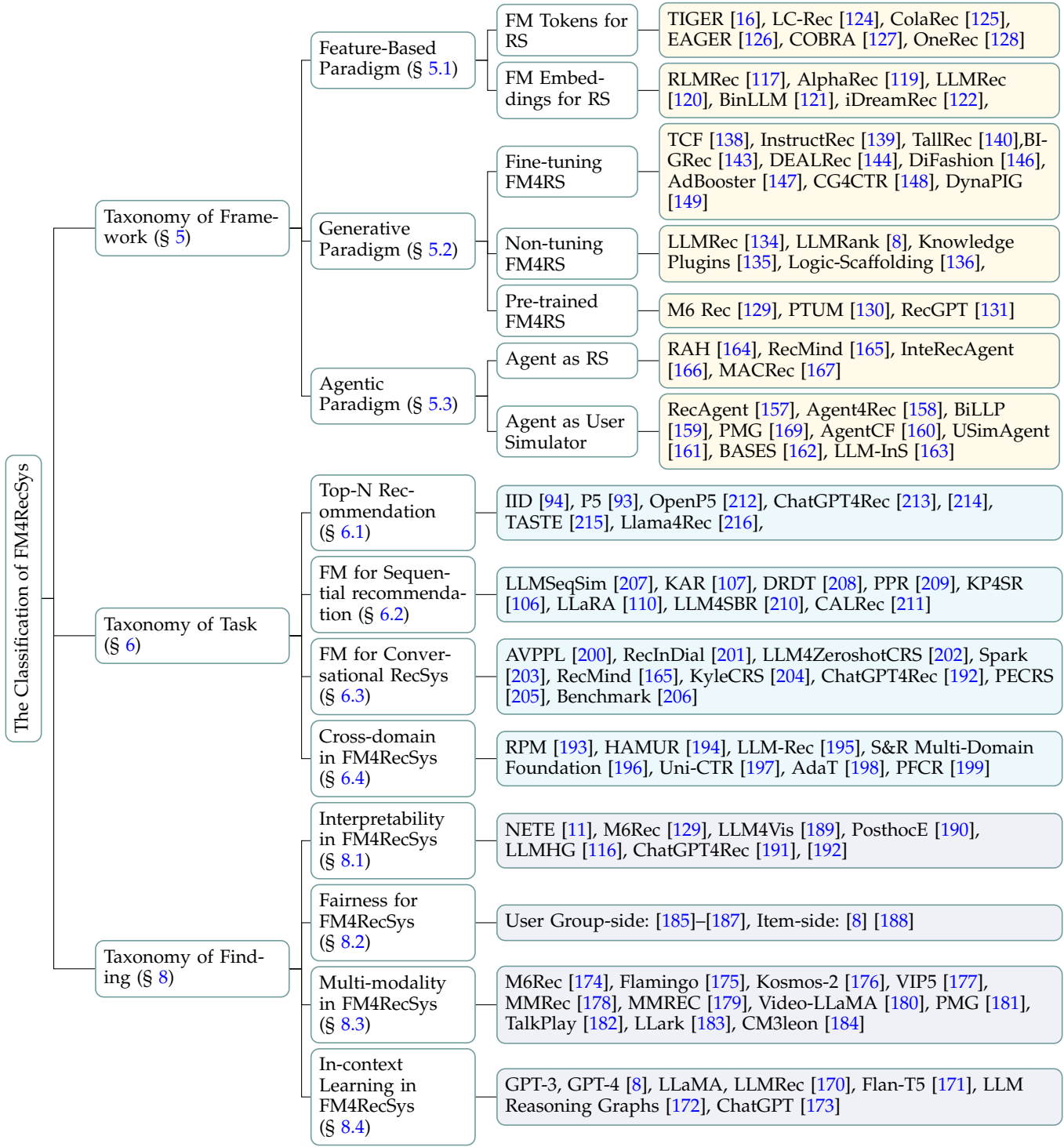

\subsection{Top-N Recommendation}
\label{subsec: top-N recommendation}

\noindent  
The Top-N recommendation task aims to generate a ranked list of the most relevant items for a user, typically representing the top-N items deemed most suitable based on their preferences \cite{anelli2022top}. 
However, if user information (including meta-information and item interaction history) is overly lengthy, it may exceed the input length capacity of foundation models. To address this, 
one way is to leverage a feature-based paradigm, using foundation models' embeddings to replace traditional user and item embeddings to execute the Top-N recommendation \cite{HarteZLKJF23, survey1}. However, the limitation of this approach is that it often ignores the rich contextual semantics and lacks the generative capabilities needed to generalize to unseen users or items. 
Another way is that foundation models use a prompt that only includes user information, asking the foundation models to directly generate recommendations for those users \cite{Xu2023,geng2022recommendation}.
In the case of multimodal and generative representation methods, the generated recommendation items can undergo similarity calculations with the multi-modal representation of ranking candidates \cite{LiuMXLYL0023}. Additionally, some approaches \cite{LiZM23,DaiSZYSXS0X23} follow practices from the NLP field. They select K-item negative samples or hard examples, feed them along with user prompts to FMs, and obtain the final rankings. However, these approaches target the idealized experimental scenario and may not be practical for real-world recommendation systems with millions of items. 
More recently, the Llama4Rec \cite{Sichun24} framework synergistically integrates both ID and text representation through data and prompt augmentation strategies and an adaptive aggregation module, resulting in significant improvements in recommendation performance.



\subsection{Sequential Recommendation}\label{sec:context}

Various FM-based approaches have been proposed to exploit their capability in the realm of context-aware recommendation. Not only can the extensive world knowledge stored in FMs serve as a rich source of background information for items \cite{HarteZLKJF23}, but also can the reasoning capability of FMs augment the next item prediction \cite{Yunjia,Wangyureasoning}. Jesse \textit{et al.}  
\cite{HarteZLKJF23} first explores three different methods of utilizing foundation models' knowledge for context-aware recommendations, based on FM semantic similarity, FM prompt fine-tuning, and BERT4Rec initialized by FM semantic embedding. After that, Artun \textit{et al.} \cite{boz2024improving} propose three orthogonal methods: LLMSeqSim, LLMSeqPrompt, and LLM2Sequential, for leveraging FMs in sequential recommendation, along with hybrid strategies that combine their strengths based on item popularity and session context. Extensive experiments across multiple datasets demonstrate that LLM-enhanced models significantly improve accuracy, diversity, and coverage, with fine-tuned GPT-3.5 notably outperforming PaLM 2 in next-item prediction tasks.
Wu \textit{et al.}~\cite{YiqingPPR} generate personalized soft prompts using user profile knowledge and employ prompt-oriented contrastive learning for effective training.
After that, Zhai \textit{et al.}~\cite{ZhaiZWL023} introduce knowledge prompt-tuning for sequential recommendations, which effectively integrates external knowledge bases with FMs, transforming structured knowledge into prompts to refine recommendations by bridging semantic gaps and reducing noise.  
Then, Liao \textit{et al.}~\cite{Jiayi} employ a hybrid approach for item representation in input prompts for FMs, combining ID-based item embeddings from traditional recommenders with textual item features, bridging the modality gap between traditional recommender systems and FMs through an adapter, and facilitating the transfer of user behavioral knowledge to the FM's input space.
The LLM4SBR~\cite{Shutong24} framework transforms session data into a bimodal form of text and behavior, leveraging large language models (LLMs) for enhanced inference and alignment.
Meanwhile, Wang \textit{et al.} \cite{Wangyureasoning} utilize the reasoning capability of Foundation Models (FMs) and introduce a collaborative in-context demonstration retrieval method, abstracting high-level user preferences and reducing noise to improve the recommendation process without the need for FM fine-tuning.
More recently, Li \textit{et al.} \cite{Yaoyiran} proposed CALRec, a contrastive-aligned generative framework for adapting LLMs to sequential recommendation. It uses two-stage fine-tuning: first, a two-tower setup with contrastive and language modeling losses on multi-domain data; then, fine-tuning on the target domain. However, CALRec struggles in cold-start scenarios, often generating item descriptions memorized from training.



\subsection{Conversational Recommendation}\label{sec:interactive}

The goal of conversational recommendation is to not only to suggest items to users over multiple rounds of interactions, but also to provide human-like responses for multiple purposes such as preference refinement, knowledgeable discussion, or recommendation justification~\cite{JannachMCC21,SunZ18}. 
The emergence of FMs has undoubtedly impacted conversational RS, especially CRS-related research.
He \textit{et al.}~\cite{HeXJSLFMKM23} presents empirical evidence that FMs, even without fine-tuning, can surpass existing conversational recommendation models in a zero-shot setting.
After that, a series of works \cite{liu2023chatgpt,spark,wang2023recmind,KyleCRS} 
adopt the role-playing prompt to guide ChatGPT/GPT-4 in simulating user interaction with conversational recommendation agents. These works augment FMs' capability through techniques such as RAG and Chain-of-Thought (CoT).
Meanwhile, several studies are built based on prior work \cite{ZhouZBZWY20} in knowledge graph-based conversational recommendation. For instance, Wang \textit{et al.} \cite{lingzhi2021} introduce a framework that integrates pre-trained language models like DialoGPT with a knowledge graph to generate dialogues and recommend items, showcasing how FM's generative capability can be utilized for conversational recommendation.
Zhang \textit{et al.} \cite{Gangyi} explores a user-centric approach, emphasizing the adaptation of FMs to users' evolving preferences through graph-based reasoning and reinforcement learning.
However, most methods rely on external knowledge graphs, require additional data labeling, and may suffer from training inefficiencies and semantic misalignment issues. In contrast, Mathieu \textit{et al.} \cite{RavautZXSL24} propose PECRS, a unified and parameter-efficient conversational recommender system (CRS). PECRS formulates CRS as a natural language processing task, directly leveraging one pre-trained FM to encode items, understand user intent, perform item recommendations, and generate dialogues. 
Recently, 
Wang \textit{et al.} \cite{wang2023rethinking} critique the current evaluation protocols for conversational RSs and introduce an FM-based user simulator approach, iEvaLM, which significantly enhances evaluation accuracy and explainability. However, FMs for conversational recommendation are still hindered by a tendency towards popularity bias and sensitivity to geographical regions \cite{lichtenberg2024large}. Meanwhile, in the context of multi-turn conversational recommendation, determining the appropriate timing for state transitions during human-RS interactions is a significant challenge. For example, deciding whether to continue the conversation with the user or make a recommendation at a given moment is crucial. A key issue that needs addressing is how to effectively model a state checker using an FM to handle these decisions.

\subsection{Cross-domain Recommendation}\label{sec:cross-domain}

In real-world scenarios, data sparsity is a pervasive issue for Collaborative Filtering (CF) recommender systems, as users rarely rate or review a broad range of items, particularly new ones. Cross-domain recommendation (CDR) tackles this by harnessing abundant data from a well-informed source domain to enhance recommendations in a data-scarce target domain. Multi-domain recommendation (MDR) extends this concept by utilizing auxiliary information across multiple domains to recommend items within those domains to specific users~\cite{ZhuW00L021}. 
However, domain conflicts remain a significant hurdle, potentially limiting the effectiveness of recommendations. The advent of foundation models that are pre-trained on extensive data across various domains and possess the cross-domain analogical reasoning ability~\cite{HuSLC23} presents a promising solution to these challenges.

HAMUR \cite{Xiaopeng} designs a domain-specific adapter to be integrated into existing models and a domain-shared hyper-network that dynamically generates adapter parameters to tackle the mutual interference and the lack of adaptability in previous models.
Tang \textit{et al.} \cite{Zuoli} discuss the application of FMs in multi-domain recommendation systems by mixing the user behavior across different domains, concatenating the title information which items into a sentence, and modeling the user behavior with a pre-trained language model, which demonstrates the effectiveness across diverse datasets. 
The S\&R (Search and Recommendation) Multi-Domain FM \cite{GongDSSLZ23} employs FMs to refine text features from queries and items, improving CTR predictions in new user or item scenarios. 
KAR~\cite{Yunjia} further leverages the power of FMs for open-world reasoning and factual knowledge extraction, and adaptation. It introduces a comprehensive three-stage process encompassing knowledge, reasoning and generation, adaptation, and subsequent utilization.
Based on the S\&R Multi-Domain FM, Uni-CTR~\cite{Zichuan} employs a unique prompting strategy to convert features into a prompt sequence that FMs can use to generate semantic representations, capturing commonalities between domains while also learning domain-specific characteristics through domain-specific networks.  
More recently, Fu \textit{et al.} \cite{Junchen} investigate the efficacy of adapter-based learning for CDR, which is designed to leverage raw item modality features, like texts and images, for making recommendations. They conduct empirical studies to benchmark existing adapters and examine key factors affecting their performance.
However, it is worth mentioning that CDR faces challenges of domain privacy leakage and ineffective knowledge transfer in existing FM4RecSys methods. To address these issues, Guo \textit{et al.}~\cite{LeiGuo24} proposed the PFCR framework, which introduces a privacy-preserving federated learning schema using local client interactions and gradient encryption. This framework models items in a universal feature space through description texts and leverages federated content representations with prompt fine-tuning strategies.

\subsection{Item and Content Generation for Recommendation}

A forward-looking application of FMs in RS is item generation – creating new content that can be recommended to the user. This goes beyond the classical remit of RS, which typically selects from existing items. However, with generative models, the line between recommending an existing item and generating a new item tailored to the user can blur. The main tasks of item and content generation as follows:
\par \textbf{Bundle Generation:} In E-commerce, bundle recommendation is crucial for increasing average order value. BundleGen \cite{zhou2023bundlegen} introduces a diffusion-based framework to generate item bundles conditioned on a seed item and user preference. Instead of retrieving co-purchased items, it learns to generate a set of compatible, stylistically coherent items using a denoising process in the item embedding space. The model iteratively refines random noise into a meaningful bundle, optimizing for both intra-bundle compatibility and personalization. Compared to autoregressive models, diffusion models offer better control over diversity and global structure, especially in high-dimensional item spaces.
\par \textbf{Playlist Generation:} In music recommendation, playlist generation is evolving from heuristic-based sequencing to fully generative paradigms. Recent work, such as MusicGen \cite{wang2023musicgen}, fine-tunes pre-trained language models like GPT-2 to autoregressively generate sequences of song IDs, conditioned on a user’s history or intent prompt (e.g., “relaxing jazz for evening”). These models treat playlist generation as a language modeling task over item IDs, allowing flexible incorporation of user intent, mood, or temporal context. By training on curated playlists, such models capture high-order dependencies and smooth transitions between tracks, outperforming retrieval-based baselines on diversity and coherence metrics.
\par \textbf{Text Content Recommendations:} In news recommendation, generative models are used to go beyond list ranking by enabling LLMs to generate summaries or narratives that align with a user’s interest profile. GNR (Generative News Recommendation) \cite{yao2022generative} adopts a two-stage pipeline: first retrieving relevant news articles, then using GPT-based models to synthesize a cohesive summary that connects the stories under a unifying theme. This approach transforms passive article lists into engaging narratives. Similarly, Prompt4NewsRec \cite{li2023promptnewsrec} uses prompt-based generation to model user interest evolution and recommend news in natural language, leveraging personalization at the prompt level.
\par  \textbf{Creative Content Personalization:} Many recent content-ware works leverage visual FMs, such as Stable Diffusion \cite{rombach2022high}, to generate personalized recommendation content. DiFashion \cite{xu2024diffusion} creates personalized outfit images from scratch based on user preferences.  AdBooster \cite{shilova2023adbooster} leverages user interest signals to personalize ad creative generation based on the Stable Diffusion. CG4CTR \cite{yang2024new} further employs Stable Diffusion to generate background images while preserving the main product details as diverse ads for different users. DynaPIG \cite{czapp2024dynamic} leverages diffusion-based FMs to create visually appealing personalized product images. 
\par \textbf{New Item Cold-Start as Generation:} Cold-start items lack behavioral data, but generative models can hallucinate metadata or simulate interactions. Acharya \textit{et al.} \cite{acharya2023leveraging} use few-shot prompting with LLMs to generate item descriptions (e.g., genre, plot, key attributes) for new movies or books based on minimal metadata (like title or category), improving recall when fed into traditional recommenders. ColdLLM \cite{huang2024coldllm} proposes a two-stage framework: a lightweight filtering model selects promising user candidates, and then GPT-4 simulates interactions (clicks, ratings) between these users and cold-start items. These synthetic logs are fed into downstream models, yielding strong gains in offline metrics and GMV in real-world A/B tests. These approaches demonstrate the generative capability of LLMs in bootstrapping cold-start items without any real interactions.



\subsection{Discussion}

In this section, we analyze the suitability of the three integration frameworks for various recommendation tasks. By comparing the inherent capabilities of Feature-Based, Generative, and Agentic paradigms, we can identify which framework is best suited for each type of task and where potential hybrid approaches might be beneficial.

For the Top-N Recommendation task, which primarily involves ranking and matching based on user preferences, the Feature-Based framework has a clear advantage. This approach focuses on extracting high-quality embeddings that capture semantic similarities, allowing for efficient and accurate ranking. While Generative methods can also operate in zero- or few-shot settings, they tend to suffer from issues of output controllability, and Agentic systems often introduce unnecessary complexity when the task does not require dynamic interaction or multi-turn feedback.

For Sequential Recommendation, the goal is to model user behavior over time and capture the evolving nature of preferences. Feature-based methods, with their emphasis on learning latent patterns from sequential data, remain effective in this area. However, Agentic frameworks show promise by integrating memory and planning capabilities to better handle long-term dependencies and adapt to changes in user behavior. Generative approaches might contribute through techniques such as chain-of-thought prompting for sequence modeling, but they usually face challenges with preserving the natural order of interactions.
For Conversational Recommendation, which inherently relies on multi-turn dialogue and context-aware interactions, both Generative and Agentic paradigms excel. The natural language generation capability of Generative models enables them to produce personalized and contextually rich responses, while Agentic systems add value by dynamically managing interactions, incorporating feedback, and adapting recommendations over the course of a conversation. This task benefits most from a combination of the two, where conversational fluency and interactive reasoning are paramount.
Cross-domain recommendation tasks require the system to generalize across diverse data sources and domains. Here, the Feature-Based framework is particularly effective when it leverages multi-modal embeddings to capture semantic similarities between different domains. Generative methods can help in bridging data sparsity by generating contextual connections, although they might lack the consistency needed for reliable domain transfer. Agentic frameworks could further refine cross-domain recommendations by dynamically integrating external data, but their added complexity may not be necessary unless adaptive, real-time adjustments are required.
Finally, for Item/Content Generation, where new content such as item descriptions, images, or creative outputs must be generated, the Generative framework is inherently best suited. Its strength lies in the ability to produce novel, high-quality content directly. Agentic approaches might complement this task through iterative planning and feedback incorporation, yet the core generation task is most effectively handled by methods that specialize in flexible, creative output.

\begin{tcolorbox}[
    title=Comparative Discussion,
    colback=white,          
    colframe=black,        
    coltitle=white,        
    colbacktitle=black,     
    fonttitle=\bfseries,    
    boxrule=0.75pt,         
    arc=2pt,                
    left=6pt, right=6pt,    
    top=6pt, bottom=6pt     
  ]
\begin{itemize}[leftmargin=*]
    \item \textbf{Top-N Recommendation:} Best served by Feature-Based methods due to robust embedding extraction and efficient ranking; Generative and Agentic frameworks add complexity that may be unnecessary for static ranking tasks.
    \item \textbf{Sequential/Conversational Recommendation:} Sequential tasks can leverage both Feature-Based and Agentic approaches for capturing temporal dependencies, while conversational tasks benefit significantly from the natural language generation of Generative models, with Agentic systems enhancing interactive adaptability.
    \item \textbf{Cross-Domain and Item Generation:} Feature-Based approaches are advantageous for cross-domain tasks by providing semantic alignment across domains, whereas Item/Content Generation is most aligned with Generative models, possibly enhanced by Agentic feedback loops for real-time refinement.
\end{itemize}
\end{tcolorbox}


\section{FM4RecSys Showcase: Empirical Comparison of FM for Sequential Recommendation}

As shown in Table \ref{tab:comparison}, we conduct a comprehensive empirical study to benchmark representative FM4RecSys frameworks across various sequential recommendation datasets. Overall, feature-based models maintain strong performance on standard benchmarks, benefiting from explicit feature engineering and effective modeling of historical interactions. For instance, models like HyperGraph-LLM and ReAT consistently achieve competitive results on the Beauty, Toys, and Sports datasets.
Generative-based models further demonstrate superior performance, especially on complex datasets with sparse user behavior signals or rich semantic information (e.g., Yelp, MIND, GoodRead). Models such as POD and GenRec leverage powerful language models (LMs) as behavior generators, achieving state-of-the-art NDCG@5 on multiple datasets, and highlighting their ability to capture complex user preference patterns.
Since most of the agentic methods take a training-free strategy.
indicating their potential for handling interactive and personalized recommendation scenarios.

This empirical comparison reveals several trends: (i) feature-based models remain strong baselines for structured recommendation tasks; (ii) generative models offer superior flexibility and generalization in diverse recommendation environments, and (iii) agentic models open up new research directions for interactive and autonomous recommender systems. These findings provide practical guidelines for the development of future FM4RecSys models and motivate further exploration of hybrid paradigms that unify the advantages of different FM families.

\begin{table*}[htbp]
\Large
\centering
\resizebox{\textwidth}{!}{
\begin{tabular}{l|l|l|c|cccccccccc}
\toprule
\rowcolor{gray!20}
\multirow{2}{*}{Reference} & \multirow{2}{*}{Framework} & \multirow{2}{*}{Backbone} & \multirow{2}{*}{NS} 
& \multicolumn{9}{c}{NDCG@5 by Dataset} \\
\cline{5-13}
& & & & Beauty & Toys & Sports & Games & CDs & Office & Yelp & MIND & GoodRead \\
\midrule 
SARSRec \cite{SARS} & - & Transformer   & 0 & 0.0249 & 0.0306 & 0.0154 & 0.0365 & - & - & 0.0100 & - & - \\
BertRec \cite{sun2019bert4rec}  & - & Bert  & 0 & 0.0124 & 0.0071 & 0.0075 &  0.0311 & - & - & 0.0033 & - & - \\
\midrule
P5 \cite{geng2022recommendation}       & Feature-based & T5      & 0 & 0.0107 & 0.0050 & 0.0041 & - & - & - & - & - & - \\
TIGER \cite{rajput2023recommender}    & Feature-based/Generative & T5      & 0 & 0.0321 & 0.0371 & 0.0181 & - & - & - & - & - & - \\
LMIndexer \cite{jin2023language} & Feature-based & T5      & 0 & 0.0262 & 0.0268 & 0.0142 & - & - & - & - & - & - \\
HyperGraph-LLM \cite{chu2024llm} & Feature-based & GPT-4  & 0 & 0.0376 & 0.0379 & - & - & - & - & - & - & - \\
ReAT \cite{cao2024aligning}     & Feature-based & T5      & 99 & 0.0382 & 0.0390 & 0.0188 & - & - & - & - & - & - \\
LC-Rec \cite{zheng2024adapting}   & Feature-based/Generative & GPT-3.5 & Random & - & - & - & 0.0560 & - & - & - & - & - \\
ONCE \cite{liu2024once}   & Feature-based & GPT-3.5 & 4 & - & - & - & - & - & - & - & 0.3872 & 0.7196 \\
EmbSum \cite{zhang2024embsum}  & Feature-based & T5      & 4 & - & - & - & - & - & - & - & 0.3675 & 0.5486 \\
LLMHD \cite{song2024large}  & Feature-based & GPT-3.5  & - & - & - & - & - & - & - & 0.0542 & - & - \\
KDA \cite{yang2024sequential}  & Feature-based & GPT-3 & 99 & - & - & - & - & - & 0.3403 & - & - & - \\
POD \cite{CIKM23-POD}      & Generative & T5   & 99 & 0.0395 & 0.0599 & 0.0396 & - & - & - & - & - & - \\
GenRec \cite{cao2024genrec}   & Generative & T5 & 0 & 0.0397 & - & 0.0332 & - & - & - & 0.0475 & - & - \\
BIGRec \cite{bao2023bi}   & Generative & Llama & - & - & - & - & 0.0189 & - & - & - & - & - \\
RDRec \cite{wang2024rdrec}    & Generative & T5   & 99 & 0.0461 & 0.0593 & 0.0408 & - & - & - & - & - & - \\
RecGPT \cite{zhang2024recgpt}   & Generative & GPT-1 & 0 & 0.0143 & 0.0355 & 0.0408 & - & - & - & 0.0107 & - & - \\
RecMind \cite{wang2023recmind}  & Agentic & GPT-3.5 & 99 & 0.0289 & - & - & - & - & - & 0.0342 & - & - \\
AgentCF \cite{AgentCF} & Agentic & GPT-3.5 & 9 & - & - & - & - & 0.4373 & 0.3589 & - & - & - \\
\bottomrule
\end{tabular}
}
\caption{Performance comparison of different FM4RecSys frameworks on sequential recommendation tasks over multiple datasets. N@5 refers to NDCG@5, and NS denotes the number of negative samples. Missing values are denoted as ’-’.}
\label{tab:comparison}
\end{table*}
\section{FM4RecSys: Explored Opportunities and Findings}\label{sec: explored opportunities and findings}

As discussed above, the emergence of FMs has opened up unprecedented opportunities for advancing RSs, fundamentally reshaping how user preferences and behaviors are modeled, predicted, and interacted with. However, integrating these powerful models into RS scenarios also presents a host of new challenges that require innovative solutions across multiple dimensions. This section outlines key research directions and recent advancements aimed at enhancing RS capabilities in the FM era. We highlight developments in dynamic temporal extrapolation, multi-modal agent intelligence, retrieval-augmented generation (RAG), explainability, life-long personalization, and system-level efficiency and scalability. Together, these emerging areas offer a comprehensive roadmap for building RSs that are more adaptive, reliable, and capable of meeting evolving user needs in complex, real-world environments.

\subsection{Explanations and Interpretation}
\label{subsec: explanations and interpretation}

A common task in enhancing the interpretability of recommendation systems is the
generation of natural language explanations~\cite{ZhangC20}. This involves
directing the recommender or external model to produce, in a sentence or a paragraph,
the reasons behind recommending a specific item to a particular user. For instance,
given a user $u$ and an item $i$, the model is tasked to generate a coherent and
understandable explanation in natural language that elucidates why item $i$ is recommended
to user $u$. A series of works use ID-based representation and leverage prompts
like ``explain to the user $u$ why item $i$ is recommended''~\cite{LiZC20}. However,
the use of IDs alone in prompts may lead to vague explanations, lacking
clarity on specific aspects of the recommendation. To address this, Cui \textit{et
al.} \cite{cui2022m6} propose to integrate item features as hint words in the
prompt, aiming to guide the model more effectively in its explanatory process.
Meanwhile, Wang \textit{et al.} \cite{LLM4Vis} introduced LLM4Vis, a ChatGPT-based
method for visualization recommendations that uses in-context learning to generate
visualizations and human-like explanations, avoiding the need for a large
dataset of examples like traditional methods. Ngoc \textit{et al.} \cite{Ngoc23}
combines embedding-based and semantic-based models to generate post-hoc
explanations, leveraging ontology-based knowledge graphs to enhance interpretability
and user trust. Then, Chu \textit{et al.} \cite{Zhixuan24} combined the
reasoning capabilities of FMs with the structural advantages of HNNs (Hypergraph
Neural Networks) to better capture and interpret individual user interests. Recently,
Liu \textit{et al.} \cite{liu2023first} leverage continuous prompt vectors instead
of discrete prompt templates. Remarkably, it is found that ChatGPT, operating
under in-context learning without fine-tuning, outperforms several traditional supervised
methods~\cite{liu2023chatgpt}.

\subsection{Fairness for FM4RecSys}\label{subsec: fairness_for_FM4RecSys}

The imperative of fairness in RS stems from its widespread use in decision-making and meeting user demands. Nonetheless, there currently exists a deficiency in comprehending the degree of fairness manifested by FMs in RSs, as well as in identifying suitable methodologies for impartially
addressing the needs of diverse user and item groups within these models~\cite{abs-2305-12090,ZhangBZWF023}.
For the user group side, Hua \textit{et al.}~\cite{abs-2305-12090} propose the
Unbiased Foundation Model for Fairness-aware Recommendation (UP5) based on Counterfactually-Fair-Prompting (CFP) techniques. After that, Zhang \textit{et al.} \cite{ZhangBZWF023} crafted metrics and a dataset that accounts for different
sensitive attributes in two recommendation scenarios: music and movies, and evaluate
 ChatGPT's fairness in RS concerning various sensitive attributes of the user
side. Recently, Deldjoo \textit{et al.} \cite{Yashar24} proposed CFaiRLLM, a comprehensive evaluation framework designed to assess and mitigate biases
in LLM-based RS by examining how recommendations vary with the inclusion
of sensitive attributes such as gender and age, underscoring the impact of
different user profile sampling strategies on fairness outcomes. However, they
only focus on age and sex, suggesting the need for future research to consider a broader array of sensitive attributes and to validate the framework across diverse
domains.

For the item side, Hou \textit{et al.}~\cite{hou2024zero} guide FMs with prompts
to formalize the recommendation task as a conditional ranking task to improve item-side fairness. Besides, Jiang \textit{et al.}~\cite{JiangBZW0F024} investigate the impact of historical user interactions and inherent semantic biases in FMs and
introduce a framework called IFairLRS, which adapts traditional fairness methods
to enhance item-side fairness in FM4RecSys without compromising recommendation accuracy. Research on non-discrimination and fairness in FM4RecSys is still in its early stages, indicating a need for further investigation. 

\subsection{Multimodal Recommender Systems}
\label{subsec: multi-modal recommender systems}

Recent years have seen the rapid advancement of MFMs that can jointly process and understand multiple modalities within a unified architecture. These foundation models have opened new frontiers in recommendation by enabling both enhanced content understanding and generative capabilities.
A key aspect of multimodal recommendation is \textit{modality fusion}, combining signals from text (e.g., item descriptions), images (e.g., product photos), audio (e.g., music or speech), and video to enrich user and item representations. Models like M6-Rec \cite{cui2022m6rec} unify recommendation tasks (retrieval, ranking, explanation, content creation) by converting multimodal inputs into a text-based format and applying large Transformer models for reasoning and generation. Similarly, Flamingo~\cite{alayrac2022flamingo} and Kosmos-2 \cite{peng2023kosmos} adopt cross-attention mechanisms that allow language models to attend to visual embeddings, enabling few-shot reasoning grounded in visual context. In domains such as fashion or home decor, incorporating image features has been shown to significantly improve item matching and ranking quality, especially in cold-start scenarios \cite{geng2023vip5}.
Frameworks such as MMRec \cite{zhou2023mmrec} and MMREC \cite{tian2024mmrec} support the integration of pre-trained visual and textual encoders into recommendation pipelines, providing flexible architectures for early fusion, co-training, and hybrid models. In the video domain, models like Video-LLaMA \cite{zhu2023videollama} represent recent breakthroughs in integrating visual frames, audio tracks, and text transcriptions into large language models to enable content-based video recommendation and captioning. These advances highlight the trend of leveraging raw multimodal content, rather than relying solely on sparse interaction data or metadata.
Beyond understanding existing content, multimodal foundation models enable \textbf{personalized content generation}, effectively blurring the boundary between recommendation and creation. In e-commerce, systems like M6-Rec\cite{cui2022m6rec} and PMG \cite{shen2024pmg} can generate personalized product images or fashion styles from scratch, conditioned on user preferences inferred from interaction histories. In the music domain, models such as TalkPlay \cite{he2024talkplay} and LLark \cite{brookes2023llark} generate textual explanations or music tags directly from audio input, enabling conversational recommendation interfaces and playlist generation with better contextual understanding.
Such generative capabilities are supported by models like CM3leon \cite{beaumont2023cm3leon}, which integrate text-to-image and image-to-text generation within a decoder-only transformer, and by retrieval-augmented models that can dynamically fetch relevant content snippets to condition output generation. These architectures open up use cases such as personalized ads, playlist artwork, video summaries, or even AI-created recommendation items tailored to each user.

Further progress in this area will involve improving the efficiency and scalability of multi-modal foundation models for real-time recommendation, developing more robust evaluation protocols for generative recommendation quality, and enhancing personalization strategies to better align multimodal generation with individual user intent. As these systems continue to mature, they are expected to support more intelligent, adaptive, and engaging recommendation experiences that unify content understanding and creation.

\subsection{In-context Learning Capability}
\label{subsec: in-context learning capability}

In-context learning, a hallmark of modern FMs, refers to the ability to perform tasks without parameter updates by conditioning on prompts that contain task descriptions and examples. Within RS, this capability manifests through zero-shot, few-shot, and reasoning-based approaches that allow LLMs to generate relevant item suggestions, often without explicit supervised training for the recommendation task. Recent work has shown that this paradigm opens new avenues for flexible, domain-agnostic, and user-friendly recommendations.

\textbf{Zero-shot and Few-shot Recommendation:} LLMs such as GPT-3, GPT-4, and LLaMA have demonstrated notable performance in zero-shot and few-shot recommendation scenarios. Researchers have formulated next-item prediction and top-N recommendation as language modeling tasks, prompting LLMs with user histories, candidate items, and instruction-based queries. Hou \textit{et al.} \cite{hou2024zero} show that GPT-4, when prompted appropriately, can act as a zero-shot ranker, exhibiting competitive ranking accuracy on benchmark datasets. Similarly, Wang \textit{et al.} \cite{wang2023zeroshot} demonstrates that a structured 3-step prompt can enable GPT-3 to outperform some traditional sequential models on MovieLens.
Few-shot strategies further enhance performance by conditioning the model on a handful of annotated recommendation examples. Benchmarks like LLMRec \cite{liu2023llmrec} reveal that while LLMs underperform on raw recommendation metrics compared to specialized models, they excel in explanation generation and preference summarization---showcasing their capacity to integrate semantics and simulate commonsense reasoning.

\textbf{Reasoning and Instruction-following:} Instruction-following LLMs can generate recommendations based on free-form user descriptions or complex goals. For instance, Zhang \textit{et al.} \cite{zhang2023instruction} reformulate recommendation as an instruction-following task, where user profiles and intents are described in natural language. Their fine-tuned Flan-T5 model not only generates relevant recommendations but also adapts flexibly across domains. Similarly, LLM Reasoning Graphs \cite{wang2023llmrg} leverage LLMs to generate logical chains linking user interests to item features, which are then used to guide downstream recommenders.
In conversational recommendation, in-context reasoning enables LLMs to process feedback iteratively. Spurlock \textit{et al.} \cite{spurlock2024chatgpt} show that ChatGPT can refine its recommendations through user feedback via reprompting, thus aligning better with user preferences. This showcases the dual strengths of LLMs in understanding nuanced natural language and adapting recommendations through multi-turn reasoning.

In-context learning enables LLMs to serve as adaptable, explainable, and domain-transferable recommenders. While their performance in traditional metrics can lag behind specialized models, their strengths in semantic reasoning, instruction adherence, and human-centric interaction make them powerful tools for next-generation recommendation systems. Ongoing research continues to explore how to better align, prompt, and integrate LLMs into broader recommendation architectures.

%



\begin{tcolorbox}[
    title=Explored Opportunities And Findings,
    colback=white,          
    colframe=black,        
    coltitle=white,        
    colbacktitle=black,     
    fonttitle=\bfseries,    
    boxrule=0.75pt,         
    arc=2pt,                
    left=6pt, right=6pt,    
    top=6pt, bottom=6pt     
  ]
\begin{itemize}[leftmargin=*]
    \item FMs have paved the way for more intuitive explanations in RSs. By integrating rich item features and leveraging advanced prompting techniques, these models now produce natural language explanations that not only clarify recommendation rationale but also boost user trust and system transparency.
    \item Emerging frameworks focused on fairness indicate promising directions for mitigating bias on both user and item sides. Although initial efforts have shown encouraging results in handling sensitive attributes, there remains a significant opportunity to broaden these techniques and ensure equitable performance across diverse application domains.
    \item The incorporation of multi-modal data and in-context learning has markedly enhanced the personalization capabilities of recommendation systems. This progress has enabled richer, more dynamic representations; however, it also highlights pressing challenges in scalability, evaluation, and computational efficiency.
\end{itemize}
\end{tcolorbox}

\section{FM4RecSys: OPEN CHALLENGES and OPPORTUNITIES}


While significant progress has been made, the field of foundation model-powered recommender systems is still in its early days. Numerous challenges remain open, and they coincide with exciting opportunities for future research. In this section, we outline some of the key open issues and potential directions, grouped into three broad categories: online deployment, enhancing recommender system capabilities, technical scalability and efficiency, and methodological improvements.



\subsection{Online Deployment}




Foundation models, particularly large language models (LLMs), have shown promising capabilities in enhancing recommendation systems through improved generalization, natural language understanding, and multi-task unification. However, deploying such models in real-time, large-scale industrial settings introduces substantial engineering and system design challenges. Several industry leaders have taken early steps toward integration. For instance, Alibaba’s M6-Rec \cite{cui2022m6rec} has been deployed as a unified foundation model across retrieval, ranking, and content generation in their e-commerce platform. It employs parameter-efficient tuning (e.g., prefix tuning) and caching strategies to meet latency constraints. Amazon leverages LLMs for cold-start enhancement by generating synthetic metadata \cite{acharya2023leveraging}, while Spotify’s LLark \cite{brookes2023llark} explores music understanding via LLMs for personalized playlist generation and natural language tagging. Despite these advancements, real-world deployment remains constrained by several bottlenecks. First, latency and throughput pose a major hurdle, as LLMs are computationally intensive and often incompatible with the sub-100ms inference requirements of online systems \cite{agrawal2024taming,ghosh2024enabling}. Techniques such as hybrid two-stage architectures (retrieval + reranking) \cite{lu2022hyrr}, early exit transformers \cite{shan2024early}, and query caching \cite{an2025hyperrag} have been proposed to mitigate this. Second, memory and cost efficiency become critical at scale, with methods like LoRA, adapter tuning \cite{geng2023vip5}, and distillation \cite{li2024recommendation} increasingly adopted to reduce the serving footprint. Third, personalization vs. generalization presents a modeling challenge, while LLMs offer strong general knowledge, encoding user-specific nuances without overfitting or memory overhead is non-trivial. Lightweight user-specific prompts or adapters offer partial solutions. Additionally, integrating FMs into existing infrastructure involves compatibility issues, as output formats and ranking scores from generative models may not align with classical CTR-based pipelines. Production deployment also requires robust evaluation protocols, since traditional metrics like NDCG may not capture the full effect of generative personalization, necessitating A/B testing and user satisfaction tracking. Lastly, concerns around explainability, trust, and model staleness remain---hallucinations from LLMs can mislead users, and large models are harder to update regularly compared to standard recommender embeddings. Despite these challenges, foundation models continue to gain traction in recommendation deployment, with hybrid, latency-aware architectures emerging as a practical compromise to unlock their potential at scale.

\subsection{Enhancing Recommender System Capabilities}

\subsubsection{Temporal Extrapolation}
Recent studies~\cite{DBLP:journals/corr/abs-2310-10196} have demonstrated that
FMs can extrapolate time series data in a zero-shot manner, achieving performance
on par with or superior to specialized models trained on specific tasks. This success is largely due to FM's ability to capture multi-modal distributions and
their inclination towards simplicity and repetition, which resonates with the repetitive
and seasonal trends commonly observed in time series data. Time series modeling,
distinct from other sequence modeling due to its variable scales, sampling rates,
and occasional data gaps, has not fully benefited from large-scale pretraining. To
address this, LLMTIME2~\cite{gruver2023large} leverages LLMs for continuous time
series prediction by encoding time series as numerical strings and treating forecasting as a next-token prediction task. This method, which transforms token distributions into continuous densities, facilitates an easy application of LLMs
to time series forecasting without the need for specialized knowledge or high computational costs, making it particularly beneficial for resource-constrained scenarios. Moreover,
by viewing user preference data as a time series sequence, these models might adeptly
adapt to long-term shifts in preferences and enhance personalization and predictive
accuracy over time, especially with the zero-shot capability of approaches like LLMTIME2, which enables a rapid adaptation to user preference change without the
need for extensive retraining.

\subsubsection{Multi-modal Agent AI for RecSys}


Multimodal Agent AI \cite{durante2024agent} is an emerging field that explores AI systems capable of perceiving and acting across various domains through a unified understanding of multimodal data. These systems leverage generative models and diverse data sources for reality-agnostic training and can operate in both physical and virtual environments. In RS field, such agents can infer user preferences, adapt to real-time feedback, and provide personalized recommendations. Notably, they have the potential to act as simulators—not only for systems but also for user behavior—enabling offline data collection and training that could reduce real-world A/B testing costs. While early works have demonstrated proof-of-concept applications in settings like route planning and medical recommendation, many capabilities remain aspirational. Key open challenges include grounding multi-modal perception in recommendation-specific tasks, ensuring alignment with real user preferences, and scaling interactive agents in practical deployment scenarios.

\subsubsection{RAG meets RecSys}

Retrieval-augmented generation (RAG) is a technique used in FMs to enhance their generative capability by integrating external data retrieval into the generative process \cite{ragsurvey}. This approach improves the accuracy, credibility, and relevance of FM outputs, particularly in knowledge-intensive tasks like information
retrieval and RS. RAG aims to address outdated knowledge, the generation of incorrect information (hallucinations), and limited domain expertise by combining the FM's internal knowledge with dynamic external knowledge bases. RAG is
suitable for enhancing the FM4RecSys, especially in modeling lifelong user behavior
sequences in real-world RS environments \cite{ragpaper}. It could potentially
ensure that the RSs remain up-to-date with continuous shifts in user preferences and trends, which is critical for precise identification and documentation
of long-term behavioral patterns. For instance, considering the input token length
restriction of FMs, RAG may be utilized to selectively extract pertinent
portions of a user's interaction history and associated external knowledge, thereby conforming to the model's input constraint. Additionally, RAG may lessen the likelihood of producing irrelevant recommendations or non-existent items (hallucinations), thereby enhancing the reliability of FM4RecSys.

\subsubsection{Explainability and Trustworthyness}
Enhancing explainability and trustworthiness in RS is always a significant
challenge, especially in the FM era. The complexity and size of FMs introduce
new hurdles in explaining FM4RecSys. There are two primary approaches to explainability
in RS: one involves generating natural language explanations for recommendations,
and the other dives into the model's internal workings. The former approach has
seen considerable exploration \cite{zhang2020explainable} in the pre-FM era, whereas the latter is less developed. There are also some works~\cite{Behnam23,Yanreasoning}
that align FMs, such as their prompts, with explicit knowledge bases like
knowledge graphs. This alignment can make the model’s decision-making process traceable
as specific paths in the knowledge graph, offering a clearer explanation. However,
these approaches are still in their preliminary phase and might be further
enhanced by techniques such as Chain/Tree of Thoughts.

\subsubsection{Personalization}


LLMs with their advanced multi-modal understanding and generation capabilities, can
process and analyze vast amounts of multi-modal data, capturing intricate user
preferences and behaviors that traditional models might overlook. This deep understanding
allows LLM-based recommender systems to provide highly tailored recommendations,
enhancing the relevance and usefulness of the suggestions. By leveraging LLMs,
recommender systems can adapt to individual user needs dynamically, considering context,
past interactions, and subtle nuances in user input, which leads to more
engaging and personalized user interactions. The ability of LLMs to generate human-like
text also plays a critical role in creating more relatable and persuasive recommendations,
ultimately driving user engagement.

Despite the remarkable advancements by LLMs, there are significant challenges in
achieving effective personalization in recommender systems. Ensuring fairness
and mitigating biases in recommendations remains a critical issue, as LLMs can inadvertently
propagate and even amplify existing biases present in the training data \cite{Isabel1,Isabel2},
leading to unfair or skewed recommendations. Moreover, there are serious
concerns regarding data privacy and security, as LLMs may inadvertently leak personal
information embedded in prompts or pre-trained data~\cite{LLM4Security,EmbeddingLeak,EmbeddingLeak2}.
Another major challenge is the static nature of existing LLMs. Given the same input,
LLMs typically produce similar outputs, whereas recommender systems need to be highly
dynamic. In the same user session, LLM behavior needs to adapt drastically based
on a few user interactions. Across sessions, LLM behavior should evolve over time
to reflect changes in user preferences and interactions~\cite{personalize1,personalize2,Chenyang}.
Addressing these challenges requires ongoing research and the development of
innovative techniques to optimize LLMs for efficiency, fairness, robustness, and
data privacy in personalization tasks.

\subsection{Technical Challenges: Efficiency and Scalability}

\subsubsection{Long-sequences in FM4RecSys}

FM4RecSys faces challenges with long input sequences due to its fixed context window limitations, impacting its efficacy in tasks requiring extensive context \cite{kitaev2019reformer,beltagy2020longformer}, like in sequential recommendations. Sequential RSs, relying on a user's comprehensive interaction history and extensive item ranking lists, often exceed the FMs' context capacity, leading to less effective recommendations. Adaptations from NLP techniques are being explored, including segmenting and summarizing inputs to fit within the context window and employing strategies like attention mechanisms and memory augmentation to enhance the focus on pertinent segments of the input. The RoPE technique \cite{su2024roformer}, with its innovative rotary position embedding, shows promise in managing long inputs and offers a potential solution for maintaining the performance of an RS despite the context window constraint of FMs.

\subsubsection{System Performance Analysis: APIs cost, Training \& Inference
Efficiency}


In the development of FM-based RSs, a critical aspect is the cost assessment,
which varies depending on the data and model selections throughout the training
and inference phases~\cite{foudnation_survey8}. The training phase incurs costs due
to the recommendation model's pretraining, fine-tuning, and algorithmic development,
where the complexity and the need for specialized engineering can drive up
expenses. During the recommendation inference stage, costs persist in the form
of system upkeep, updates, and computational demands of API-driven service
provision. For instance, systems like OpenAI's GPT-3/4~\cite{brown2020language,GPT4report}
have costs associated with API usage and token interactions, escalating with more
intricate or extensive usage. Furthermore, the incorporation of RAG tools can further
elevate expenses by extending prompt lengths and, consequently, the number of tokens
processed, leading to higher API fees. Additionally, customization through fine-tuning
also adds to the overall expenses.


                    

\begin{table*}[ht]
\centering
\setlength{\tabcolsep}{12pt} 
\begin{tabular}{p{5cm}p{5cm}p{5cm}} 
\toprule
\rowcolor{gray!20}
\textbf{Cost and Efficiency} & \textbf{Possible Methods} & \textbf{References} \\ \midrule

\multirow{4}{*}{\textit{Training Cost}} 
&  Data Selection for Pre-training &  \cite{GlassGCFPBGS20,DataSelection2} \\
&  Data Selection for Fine-tuning &  \cite{Chunting,Yihan,efficientFT} \\
& Parameter-efficient Fine-tuning    & \cite{Junchen} \\
& Model Distillation &  \cite{distillation24} \\
\midrule                              
\multirow{2}{*}{\textit{Inference Latency}} & Embedding Caching & \cite{hou2023learning,HarteZLKJF23} \\
                               & Lightweight FMs   &   \cite{ChenFC0ZWC20}, \cite{Xiaoqi}, \cite{Jilin} \\
\midrule                            
\multirow{3}{*}{\textit{API Cost}} & Data Selection   &  \cite{HaoChen} \\
                               & Prompt Selection    &\cite{Jessewu} \\  & Adaptive RAG     &\cite{Mallen,huang2024learn} \\              
\bottomrule
\end{tabular}
\vspace{0.5cm} 
\caption{An organization of representative methods for reducing the cost and improving the efficiency for FM4RecSys.}
\label{cost_table}
\end{table*}

Addressing efficiency in FM4RecSys is a practical challenge with direct implications
for system performance and resource utilization. Referencing Table~\ref{cost_table},
we outline targeted solutions:

\textbf{Training Cost Reduction:} For pre-training or fine-tuning Foundation Models within recommender systems, it is necessary to carefully select the most
informative and diverse data so that the model can efficiently capture essential
user-item interaction patterns and features and accelerate the learning process~\cite{GlassGCFPBGS20,DataSelection2,efficientFT}.
Meanwhile, model distillation~\cite{KD} offers an alternative approach to
reducing fine-tuning costs. Wang \textit{et al.}~\cite{distillation24} propose
the Step-by-step Knowledge Distillation Framework for Recommendation (SLIM), which
leverages the reasoning capabilities of FMs in a resource-efficient manner. By
using Chain-of-Thought (CoT) prompting and rationale distillation, SLIM enables smaller
models to perform effective and meaningful sequential recommendations at lower costs.
In addition, employing techniques~\cite{Junchen} like LoRA~\cite{lora} and LoftQ~\cite{Yixiao}
for fine-tuning can help in managing memory usage and reducing training time.

\textbf{Inference Latency Reduction:} The computational demand for FM inference is
notable. Strategies such as employing pre-computed embedding caches~\cite{hou2023learning,HarteZLKJF23}
(e.g., VQ-Rec or LLM4Seq) can offer some relief by speeding up inference. Similarly,
efforts to compact the model size through distillation~\cite{Xiaoqi}, pruning~\cite{ChenFC0ZWC20},
and quantization~\cite{Jilin} can lead to improvements in memory cost and
inference speed. Recently, Kaur \textit{et al.} \cite{kaur2024efficient} has proposed a hybrid
task allocation strategy that leverages the strengths of both LLMs and traditional
RSs to minimize inference latency. This strategy involves two criteria for
categorizing users into two groups: strong users, who continue to receive
recommendations from traditional RSs, and weak users, who are shifted to LLM
recommendations if LLMs demonstrate superior performance.

\textbf{API Cost Reduction:} In FM-based API recommender systems, efficient data
selection can enhance the fine-tuning efficiency by using a selected set of data
points~\cite{HaoChen}. Additionally, refining prompt engineering with methods like
prompt generation or compression~\cite{Jessewu} may lead to more efficient processing
of FM inputs by making prompts more concise or better tailored, though the gains
should be considered within realistic expectations. Besides, utilizing RAG to
enhance API-based RSs can result in an additional context length, especially when
retrieving longer item descriptions as prompt inputs. Therefore, adopting adaptive
RAG~\cite{Mallen} is also an effective method to reduce API costs in that case.

\subsection{Methodological Improvements: Evaluation and Development}

\subsubsection{Benchmarks and Evaluation Metrics}

Liu \textit{et al.}~\cite{benchmarking} benchmark four state-of-the-art LLMs on five recommendation system tasks, using both quantitative
and qualitative methods. However, they only focus on specific LLMs like ChatGPT and
ChatGLM, limiting experiments to the Amazon Beauty dataset due to the high
computational cost. Thus, because of the domain-specific nature of
recommendation systems, there is a need for more datasets, recommendation tasks,
and evaluation metrics to create a more unified benchmark. Moreover, for multi-modal
and personalized agents FMs, devising new benchmarks and evaluation metrics
specifically for recommendation scenarios is essential. Finally, data pollution
presents a significant challenge. Ensuring that the evaluation data has not been
inadvertently used during FM training is difficult, making it a crucial issue to
address in the current setting.

In summary, to comprehensively evaluate and enhance the performance of FM-based recommendation
systems, a holistic and diversified benchmark is essential. Such a benchmark should
include a variety of datasets, diverse recommendation tasks, and metrics that are
adaptable to different models.

\subsubsection{Causality in FM4RecSys}


The rise of FMs has notably expanded the potential for exploring causality in
recommender systems. Equipped with vast knowledge bases and sophisticated architectures,
FMs offer unique opportunities for causal discovery and analysis, making them a major
focus in current RS research efforts. Although FMs are talented at predicting
user preferences, interpreting the causal relationships behind these preferences
is increasingly crucial. The pursuit of causal inference in recommender systems
seeks to provide recommendations that are not only transparent and reliable but
also more interpretable. A key open question is how to effectively apply causal methods
to address pressing concerns regarding bias and fairness within these systems.
These areas are critically important and represent key directions for future
research, demanding extensive exploration to uncover and address the underlying complexities
of causality in recommendation systems.

\subsubsection{Safety and Trust in FM4RecSys}

We draw on a concise discussion of how the impressive understanding and
generation power of FMs act as a dual-edged weapon in the context of FM4RecSys.

\par {\it From a perspective of safety}, FMs are vulnerable to red teaming attacks,
where malicious actors craft poison prompts to manipulate the models into
producing undesirable content. This content can range from fraudulent or racist material
to misinformation or content inappropriate for younger audiences, potentially
causing significant societal harm and putting users at risk~\cite{Boyi23, Jinghaoattack24}.
Thus, in the context of FM4RecSys especially when employing conversational
interfaces, aligning FMs with human values becomes crucial. This alignment
involves gathering relevant negative data and employing supervised fine-tuning
techniques such as online and offline human preference training~\cite{YufeiAlign,Guohaihumanalign}.
These methods can help in refining the models to adhere more closely to human instructions
and expectations, ensuring the generative contents made by FM4RecSys are safe, reliable,
and ethically sound.

\par {\it From a perspective of privacy,} if FMs are trained directly on a large
amount of sensitive user interaction data, it might be possible for third parties
to use methods like prompt injection to access specific user interaction histories,
thereby constructing user profiles. In that sense, the incorporation of
approaches such as federated learning~\cite{YangYu} and machine unlearning~\cite{Chen0ZD22,Unlearning24}
into FM4RecSys represents a promising direction for the future.


\section{Conclusion}

In this paper, we have furnished a thorough review of FM4RecSys, providing detailed comparisons and highlighting future research paths. We proposed a classification scheme for organizing and clustering existing publications and discussed the advantages and disadvantages of using foundation models for recommendation tasks. In addition, we detailed some of the most pressing open problems and promising future extensions. We hope this survey provides an overview of the challenges, recent progress, open questions, and opportunities in foundation models to the recsys research community.

\bibliographystyle{IEEEtran}
\bibliography{sample-base}

\newpage

 




\vfill

\end{document}